\batchmode
\makeatletter
\def\input@path{{"C:/Trabajo laptop/Mis articulos/Out-of-sequence PMBM/Accepted/"}}
\makeatother
\documentclass[english]{IEEEtran}
\usepackage[T1]{fontenc}
\usepackage[latin9]{inputenc}
\usepackage{array}
\usepackage{multirow}
\usepackage{amsmath}
\usepackage{amsthm}
\usepackage{amssymb}
\usepackage{graphicx}
\PassOptionsToPackage{normalem}{ulem}
\usepackage{ulem}

\makeatletter

\providecommand{\tabularnewline}{\\}

\theoremstyle{plain}
\newtheorem{thm}{\protect\theoremname}
\theoremstyle{definition}
\newtheorem{example}[thm]{\protect\examplename}
\theoremstyle{plain}
\newtheorem{prop}[thm]{\protect\propositionname}
\theoremstyle{plain}
\newtheorem{lem}[thm]{\protect\lemmaname}

\usepackage{cite}
\allowdisplaybreaks 

\usepackage[margin=8pt,font=footnotesize]{caption}
\usepackage{algorithm}
\usepackage{algpseudocode}
\usepackage{amsmath}  

\makeatother

\usepackage{babel}
\providecommand{\examplename}{Example}
\providecommand{\lemmaname}{Lemma}
\providecommand{\propositionname}{Proposition}
\providecommand{\theoremname}{Theorem}

\begin{document}
\title{Continuous-discrete multiple target tracking with out-of-sequence
measurements }
\author{Ángel F. García-Fernández, Wei Yi\thanks{A. F. García-Fernández is with the Department of Electrical Engineering and Electronics, University of Liverpool, Liverpool L69 3GJ, United Kingdom, and also with the ARIES Research Centre, Universidad Antonio de Nebrija,  Madrid, Spain (angel.garcia-fernandez@liverpool.ac.uk).  
Wei Yi is with the School of Information and Communication Engineering, the University of Electronic Science and Technology of China, China (kussoyi@gmail.com).} }
\maketitle
\begin{abstract}
This paper derives the optimal Bayesian processing of an out-of-sequence
(OOS) set of measurements in continuous-time for multiple target tracking.
We consider a multi-target system modelled in continuous time that
is discretised at the time steps when we receive the measurements,
which are distributed according to the standard point target model.
All information about this system at the sampled time steps is provided
by the posterior density on the set of all trajectories. This density
can be computed via the continuous-discrete trajectory Poisson multi-Bernoulli
mixture (TPMBM) filter. When we receive an OOS measurement, the optimal
Bayesian processing performs a retrodiction step that adds trajectory
information at the OOS measurement time stamp followed by an update
step. After the OOS measurement update, the posterior remains in TPMBM
form. We also provide a computationally lighter alternative based
on a trajectory Poisson multi-Bernoulli filter. The effectiveness
of the two approaches to handle OOS measurements is evaluated via
simulations.
\end{abstract}


\section{Introduction}

Multiple target tracking (MTT) systems are ubiquitous in many applications
ranging from air-traffic control to driving assistance systems \cite{Challa_book11,Li93,Fortin15}.
In MTT, there are an unknown number of targets that may appear, move
and disappear from a scene of interest, and the objective is to infer
their trajectories based on noisy sensor measurements. 

In multi-sensor tracking systems, these measurements are usually obtained
in scans and are sent to a processing center. Due to different time
delays in transmission, a measurement scan may be received out-of-sequence
(OOS). That is, the processing center has already processed some up-to-date
information, and receives sensor information obtained at a past time.
To use all available sensor information and improve tracking performance,
it is of interest to process these OOS measurements in a computationally
efficient manner, i.e., without having to reprocess previously received
measurements \cite{Koch11}. 

Optimal algorithms to process an OOS measurement for a single target
in linear Gaussian systems were provided in \cite{Bar-Shalom02,Bar-Shalom04,Zhang05},
and for nonlinear/non-Gaussian systems in \cite{Zhang12}. Processing
an OOS measurements can be done with a retrodiction step, which obtains
target information at the time stamp of the OOS measurement, and a
measurement update. This approach was extended in \cite{Orton05,Koch11,Govaers14}
to consider the posterior of a single trajectory, and in \cite{Shen09},
to include multiple OOS measurements. OOS measurement processing algorithms
for a fixed and known number of targets, or with external track initiation
and termination, are provided in \cite{Orton02b,Zhang03,Maskell06}
and \cite{Mallick02,Chan08}, respectively. An approximate algorithm
for processing OOS measurements within a probabilistic hypothesis
density filter is provided in \cite{Yang20}.

In this paper, we derive the exact Bayesian update with an OOS set
of measurements, with a continuous-time time stamp, for multi-target
systems. In this setting, at the OOS measurement time, we have to
account for target appearances and disappearances in continuous time,
including the possible existence of targets that did not exist at
previously sampled time steps.  

In order to explain the processing of an OOS measurement, we first
review how to process the in-sequence measurements in a Bayesian manner.
We consider a continuous-time multi-target system \cite{Angel20,Coraluppi14,Angel20_f}
in which target appearance and disappearance are given by an $\mathrm{M}/\mathrm{M}/\infty$
queuing system \cite{Kleinrock_book76} and single-target dynamics
are modelled by a stochastic differential equation (SDE) \cite{Sarkka_book19}.
This multi-target system can be discretised at the time steps when
we receive in-sequence measurements to obtain a standard multi-target
dynamic model \cite{Mahler_book14}, which consists of a time-dependent
probability of survival, single-target transition density and a Poisson
point process (PPP) birth model \cite{Angel20}. 

All information on the set of all (sampled) trajectories, i.e., trajectories
that have been discretised at the time steps when we receive the measurements,
is contained in its posterior density \cite{Angel20_b}. For in-sequence
measurements, the posterior is a Poisson multi-Bernoulli mixture (PMBM)
that can be calculated by the trajectory Poisson multi-Bernoulli mixture
(TPMBM) filter \cite{Granstrom18,Granstrom19_prov2,Angel20_e} with
the resulting discretised multi-target dynamic model. The TPMBM filter
is an extension of the PMBM filter \cite{Williams15b,Angel18_b} for
sets of targets to sets of trajectories. When the system is modelled
in continuous time, we refer to the TPMBM filter as the continuous-discrete
TPMBM (CD-TPMBM) filter. The TPMBM and PMBM filters are state-of-the-art
multiple hypothesis tracking algorithms \cite{Reid79}, with a Bayesian
birth model and an efficient representation of the posterior via probabilistic
target existence \cite{Williams15b,Brekke17,Brekke18} and a PPP intensity
to keep undetected target information, which is important, for example,
in search-and-track operations \cite{Bostrom-Rost21_early}.

\begin{figure}
\begin{centering}
\includegraphics[scale=0.6]{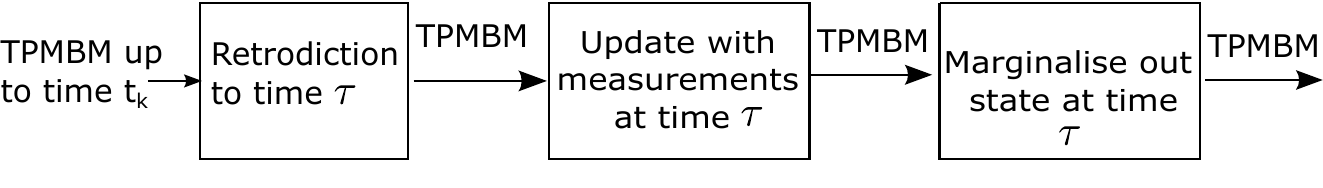}
\par\end{centering}
\caption{\label{fig:Diagram-OOS}Diagram of the update with a set of OOS measurements
at time $\tau$. The posterior over the set of all (sampled) trajectories
up to the current time step $t_{k}$ is a PMBM. To process the OOS
measurements, we perform a retrodiction step, which yields a PMBM
density that includes trajectory information at time $\tau$. The
update of a TPMBM density results in another TPMBM density. We marginalise
out the trajectory information at time $\tau$ to only keep trajectory
information on the in-sequence sampled times, which yields a PMBM
density.}
\end{figure}

The first contribution of this paper is that we derive the Bayesian
processing of an OOS set of measurements in an MTT system by applying
a retrodiction step followed by a measurement update. The retrodiction
step takes into account continuous-time target appearances, dynamics
and disappearances. This step adds state information at the OOS measurement
time for the previously sampled trajectories and new trajectories
that were not discretised at the in-sequence sampling times. Importantly,
we show that the posterior keeps the PMBM form after the retrodiction
and update steps. In order to only keep trajectory information at
in-sequence sampling times, we then marginalise out trajectory information
at OOS measurement time, which also keeps the PMBM form \cite{Granstrom20b},
see Figure \ref{fig:Diagram-OOS}.

The second contribution of this paper is to derive the Gaussian implementation
of the OOS measurement processing when the SDE corresponds to the
Wiener velocity model \cite{Sarkka_book19}. To enable a Gaussian
implementation of the CD-TPMBM filter, we first obtain the best Gaussian
PPP approximation to the birth model by minimising the Kullback-Leibler
divergence (KLD) \cite{Angel20}. The resulting discretised model,
along with linear/Gaussian measurement models with constant probability
of detection, directly allows us to implement the CD-TPMBM filter
in its Gaussian form \cite{Granstrom18,Angel20_e}. To carry out the
OOS measurement processing for the Gaussian CD-TPMBM filter, we also
require a KLD minimisation to account for new trajectories at the
OOS time. 

The obtained OOS measurement update can also be used with the continuous-discrete
(track-oriented) trajectory Poisson multi-Bernoulli (CD-TPMB) filter,
which is an approximation to the CD-TPMBM filter that only has one
mixture component \cite{Angel20_e,Williams15b,Frohle20}. The (track-oriented)
Poisson multi-Bernoulli filter is a variant of the joint integrated
probabilistic data association filter \cite{Challa_book11} that accounts
for the influence of undetected targets in the association events
\cite[Sec. IV.A]{Williams15b}. Finally, we evaluate the benefits
of OOS measurement processing for both the CD-TPMBM and CD-TPMB filters
via simulations. 

The rest of the paper is organised as follows. Section \ref{sec:Background}
provides an overview on the considered models. Section \ref{sec:Continuous-discrete-TPMBM-filter}
explains the continuous-discrete models and the CD-TPMBM filter. The
update of the CD-TPMBM filter with an OOS measurement is addressed
in Section \ref{sec:TPMBM-update-OoS}. Section \ref{sec:OOS-Gaussian}
provides the Gaussian implementation of the OOS measurement update.
Simulation results are provided in Section \ref{sec:Simulations}.
Finally, conclusions are drawn in Section \ref{sec:Conclusions}. 

\section{Background\label{sec:Background}}

This section provides a general background on the models used to solve
the problem of continuous-discrete multiple target tracking with in-sequence
measurements. The main notation of the paper is summarised in Table
\ref{tab:Notation}.

\begin{table}

\caption{\label{tab:Notation}Notation}

\rule[0.5ex]{1\columnwidth}{1pt}
\begin{itemize}
\item $\mathbf{x}_{k}$: set of targets at time step $k$, $x\in\mathbf{x}_{k}$
is a target state.
\item $\mathbf{X}_{k}$: set of all sampled trajectories up to time step
$k$.
\item $X=\left(\beta,x^{1:\nu}\right)\in\mathbf{X}_{k}$: a trajectory state,
with start time step $\beta$, length $\nu$ and states $\left(x^{1},...,x^{\upsilon}\right)$
(sampled at the in-sequence sampling times).
\item $\mathbf{Y}_{k}$: set of all sampled trajectories up to time step
$k$, including information at OOS time $\tau$.
\item $\left(u,\beta,x^{1:\nu}\right)\in\mathbf{Y}_{k}$: a trajectory with
information at OOS time $\tau$. 
\begin{itemize}
\item $u=1$: trajectory exists at OOS time with state $x^{\nu}$.
\item $u=1$, $\beta=-1$, $\nu=1$: OOS new trajectory (it was not sampled
at the in-sequence sampling times, e.g. the blue one in Fig. \ref{fig:Illustration-trajectories}).
\item $u=0$: trajectory does not exist at OOS time.
\end{itemize}
\item $f_{k|k'}\left(\cdot\right)$: density of $\mathbf{X}_{k}$ given
measurements up to time step $k'$.
\item $f_{\tau,k|k}\left(\cdot\right)$: density of $\mathbf{Y}_{k}$ given
measurements up to time step $k$, but not at time $\tau$. 
\item $f_{\tau,k|,\tau,k}\left(\cdot\right)$: density of $\mathbf{Y}_{k}$
given measurements up to time step $k$, including time $\tau$. 
\item $\lambda$: rate of appearance of new targets.
\item $\mu$: rate of the exponentially distributed life span of a target.
\item $g_{\left(\Delta t_{k}\right)}\left(\cdot\left|x\right.\right)$:
single target transition density from state $x$ with a time interval
$\Delta t_{k}$.
\end{itemize}
\rule[0.5ex]{1\columnwidth}{1pt}
\end{table}

\subsection{Sets of targets\label{subsec:Sets-of-targets}}

The multi-target state at time $t$, where $t\in\left[0,\infty\right)$,
is the set $\mathbf{x}\left(t\right)\in\mathcal{F}\left(\mathbb{R}^{n_{x}}\right)$,
where $\mathbb{R}^{n_{x}}$ is the single-target space, and $\mathcal{F}\left(\mathbb{R}^{n_{x}}\right)$
is the space of all finite subsets of $\mathbb{R}^{n_{x}}$. Targets
move independently with a continuous time model and, at any time $t$,
targets may be added or removed from $\mathbf{x}\left(t\right)$.
These models will be explained in Section \ref{subsec:Continuous-time-model}.

At time step $k\in\mathbb{N}\cup\left\{ 0\right\} $, which corresponds
to a time $t_{k}$, we take noisy measurements from the multi-target
state $\mathbf{x}_{k}=\mathbf{x}\left(t_{k}\right)$. These measurements
are in-sequence, which means that  $t_{k}>t_{k-1}$. At time step
$k$, the set $\mathbf{z}_{k}\in\mathcal{F}\left(\mathbb{R}^{n_{z}}\right)$
of measurements follows the standard point target measurement model
\cite{Mahler_book14}. That is, the set $\mathbf{z}_{k}$ is the union
of the set of target-generated measurements and the set of clutter
measurements. Given $\mathbf{x}_{k}$, each target $x\in\mathbf{x}_{k}$
is detected with probability $p^{D}\left(x\right)$ and generates
a measurement with conditional density $l\left(\cdot|x\right)$, or
missed with probability $1-p^{D}\left(x\right)$. The clutter process
is an independent PPP with intensity $\lambda^{C}\left(\cdot\right)$.

The posterior density of $\mathbf{x}_{k}$ given the sequence $\mathbf{z}_{1:k}=\left(\mathbf{z}_{1},...,\mathbf{z}_{k}\right)$
of measurements is a PMBM density that can be computed via the prediction
and update equations with continuous-discrete dynamic models \cite{Williams15b,Angel20},
which will be explained in Section \ref{subsec:Continuous-discrete-model}.

\subsection{Sets of sampled trajectories}

\begin{figure}
\begin{centering}
\includegraphics[scale=0.6]{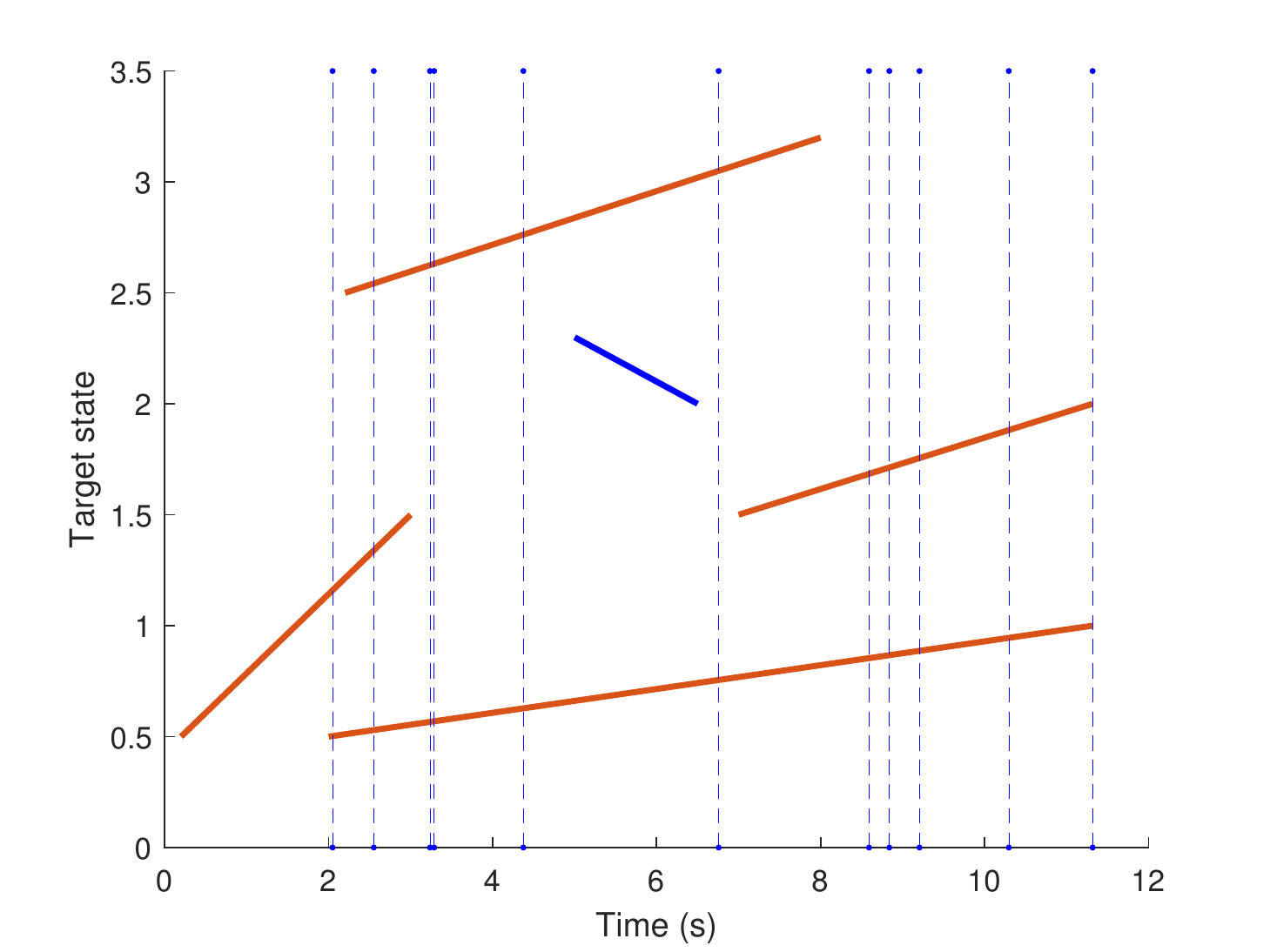}
\par\end{centering}
\caption{\label{fig:Illustration-trajectories}Illustration of a set of one-dimensional
trajectories in continuous time and its discretisation. The vertical
dashed lines indicate the times at which measurements have been taken,
which are used to discretise the trajectories. The set $\mathbf{X}_{k}$
of sampled trajectories is shown in red (the trajectories exist at
least at one of the sampled times). The blue trajectory has not been
sampled and does not belong to $\mathbf{X}_{k}$. }
\end{figure}

In order to include target trajectory information in the filter, we
consider target trajectories up to the current time $t_{k}$ sampled
at the times when the in-sequence measurements are taken. Specifically,
a trajectory is characterised by its initial time step $\beta\in\left\{ 0,1,...,k\right\} $,
its length $\upsilon$ (number of time steps that the trajectory has
been present) and, its sequence $x^{1:\nu}=\left(x^{1},...,x^{\upsilon}\right)$
of target states from time step $\beta$ to time step $\beta+\nu-1$.
A trajectory up to time step $k$ is a variable $X=\left(\beta,x^{1:\nu}\right)$,
where $\left(\beta,\nu\right)$ belongs to the set $I_{(k)}=\left\{ \left(\beta,\nu\right):0\leq\beta\leq k\,\mathrm{and}\,1\leq\nu\leq k-\beta+1\right\} $,
which ensures that the beginning and end of the trajectory belong
to the considered time window. The single-trajectory space up to time
step $k$ is $T_{\left(k\right)}=\uplus_{\left(\beta,\nu\right)\in I_{(k)}}\left\{ \beta\right\} \times\mathbb{R}^{\nu n_{x}}$,
where $\uplus$ stands for disjoint union, which is used to highlight
that the sets are disjoint. The set of (sampled) trajectories up to
time step $k$ is denoted by $\mathbf{X}_{k}\in\mathcal{F}\left(T_{\left(k\right)}\right)$.
\begin{example}
\label{exa:Set_trajectories}We consider one-dimensional targets and
the five trajectories in continuous time shown in Figure \ref{fig:Illustration-trajectories}.
We have received measurements at the times indicated by the vertical
dashed lines. The continuous trajectories are discretised at these
time steps to obtain a set $\mathbf{X}_{k}$ of (sampled) trajectories.
For example, the trajectory that appears first is (approximately)
represented in discretised form as $\left(1,\left(1.16,1.34\right)\right)$.
This means that it was born with the first round of measurements with
a state 1.16, has a duration of two time steps, and has a state 1.34
at time step two. The discretised version of the rest of the trajectories
is obtained analogously. The blue trajectory does not belong to $\mathbf{X}_{k}$,
as it appeared and disappeared in between sampling times. These types
of unobserved trajectories will play an important role in OOS measurement
updates, see Section \ref{sec:TPMBM-update-OoS}. $\diamondsuit$
\end{example}

Similarly to integrals on a single-target space $\mathbb{R}^{n_{x}}$,
we can define integrals on the single-trajectory space $T_{\left(k\right)}$.
For a real-valued function $\pi\left(\cdot\right)$ on the single-trajectory
space, its integral is \cite{Angel20_b}
\begin{align}
\int\pi\left(X\right)dX & =\sum_{\left(\beta,\nu\right)\in I_{(k)}}\int\pi\left(\beta,x^{1:\nu}\right)dx^{1:\nu}.\label{eq:single_trajectory_integral}
\end{align}
This integral sums over all possible start times and lengths, and
integrates the sequence of states. Integral (\ref{eq:single_trajectory_integral})
is the basis for the set integral on trajectories \cite{Angel20_b}.

\section{Continuous-discrete trajectory PMBM filter\label{sec:Continuous-discrete-TPMBM-filter}}

This section describes the CD-TPMBM filter for in-sequence measurements.
Before this, Sections \ref{subsec:Continuous-time-model} and \ref{subsec:Continuous-discrete-model}
review the continuous and continuous-discrete multi-target models.

Following \cite{Angel20}, we use the terms target appearance and
disappearance for the continuous time process, and target birth and
death for the discretised process. A target appearance may not imply
target birth, as the target may appear and disappear in between two
sampling times, see Figure \ref{fig:Illustration-trajectories}.

\subsection{Continuous time multi-target model\label{subsec:Continuous-time-model}}

The continuous time multi-target model has the following characteristics
\cite{Angel20}. A Poisson process (in time) with rate $\lambda$
models the times of target appearances \cite{Kleinrock_book76}. The
life span of a target is independent and exponentially distributed
with rate $\mu$. These two properties define an $\mathrm{M}/\mathrm{M}/\infty$
queuing system \cite{Kleinrock_book76} for the evolution of the number
of targets across time. 

The distribution of a target state at the time of appearance is an
independent Gaussian with mean $\overline{x}_{a}$ and covariance
matrix $P_{a}$. Targets move independently following an SDE \cite{Sarkka_book19}
\begin{align}
dx\left(t\right) & =Ax\left(t\right)dt+Ld\varpi\left(t\right)\label{eq:target_SDE}
\end{align}
where $x\left(t\right)\in\mathbb{R}^{n_{x}}$ is the target state
at time $t$, $A\in\mathbb{R}^{n_{x}\times n_{x}}$ and $L\in\mathbb{R}^{n_{x}\times n_{\beta}}$
are matrices, $dx\left(t\right)$ is the differential of $x\left(t\right)$,
and $\varpi\left(t\right)\in\mathbb{R}^{n_{\varpi}}$ is a Brownian
motion with diffusion matrix $Q_{\varpi}$.

\subsection{Continuous-discrete multi-target model\label{subsec:Continuous-discrete-model}}

The continuous time model in Section \ref{subsec:Continuous-time-model}
is discretised at the times when we receive in-sequence measurements.
The resulting discretised model results in a (time-dependent) standard
multi-target dynamic model \cite{Mahler_book14}, in which targets
evolve independently and target birth is also independent. In particular,
given $\mathbf{x}_{k-1}$, each $x\in\mathbf{x}_{k-1}$ survives to
time step $k$ with a probability of survival
\begin{align}
p_{k}^{S} & =e^{-\mu\Delta t_{k}},\label{eq:Probability_survival}
\end{align}
where $\Delta t_{k}=t_{k}-t_{k-1}$, and moves to a new state with
transition density $g_{\left(\Delta t_{k}\right)}\left(\cdot\left|x\right.\right)$
\cite{Sarkka_book19}
\begin{align}
g_{\left(\Delta t_{k}\right)}\left(x\left(t_{k}\right)\left|x\left(t_{k-1}\right)\right.\right) & =\mathcal{N}\left(x\left(t_{k}\right);F_{\left(\Delta t_{k}\right)}x\left(t_{k-1}\right),Q_{\left(\Delta t_{k}\right)}\right)\label{eq:single_target_transition_density}\\
F_{\left(\Delta t_{k}\right)} & =\exp\left(A\Delta t_{k}\right)\label{eq:discretised_F}\\
Q_{\left(\Delta t_{k}\right)} & =\int_{0}^{\Delta t_{k}}\exp\left(A\left(\Delta t_{k}-\xi\right)\right)LQ_{\varpi}L^{T}\nonumber \\
 & \quad\times\exp\left(A\left(\Delta t_{k}-\xi\right)\right)^{T}d\xi\label{eq:discretised_Q}
\end{align}
where superscript $T$ denotes transpose, $\exp\left(A\right)$ denotes
the matrix exponential of $A$ and $\mathcal{N}\left(x;\overline{x},Q\right)$
denotes a Gaussian density with mean $\overline{x}$ and covariance
matrix $Q$ evaluated at $x$. 

Targets are born according to a PPP with intensity
\begin{align}
\lambda_{k}^{B}\left(x_{k}\right) & =\frac{\lambda}{\mu}\left(1-e^{-\mu\Delta t_{k}}\right)\int_{0}^{\Delta t_{k}}p\left(x_{k}\left|t\right.\right)p_{\left(\Delta t_{k}\right)}\left(t\right)dt\label{eq:intensity_birth}\\
p\left(x_{k}\left|t\right.\right) & =\mathcal{N}\left(x_{k};F_{\left(t\right)}\overline{x}_{a},F_{\left(t\right)}P_{a}F_{\left(t\right)}^{T}+Q_{\left(t\right)}\right)\label{eq:single_target_density_given_t}\\
p_{\left(\Delta t_{k}\right)}\left(t\right) & =\frac{\mu}{1-e^{-\mu\Delta t_{k}}}e^{-\mu t}\chi_{\left[0,\Delta t_{k}\right)}\left(t\right)\label{eq:density_time_lag}
\end{align}
where  $\chi_{\left[0,\Delta t_{k}\right)}\left(t\right)=1$ if $t\in\left[0,\Delta t_{k}\right)$
and $\chi_{\left[0,\Delta t_{k}\right)}\left(t\right)=0$ otherwise.
The quantity $\frac{\lambda}{\mu}\left(1-e^{-\mu\Delta t_{k}}\right)$
is the expected number of targets that are born at time step $k$,
i.e., targets that appeared between times $t_{k-1}$ and $t_{k}$
and are still alive at time $t_{k}$ \cite{Angel20,Kulkarni_book16}.
For example, the blue trajectory in Figure \ref{fig:Illustration-trajectories}
is not considered in the birth model as it has not been sampled. Eq.
(\ref{eq:density_time_lag}) is a truncated exponential density with
parameter $\mu$ in the interval $\left[0,\Delta t_{k}\right)$ and
represents the density of the time lag $t$ of new born targets. That
is, if a target appears at time step $t_{k}-t$ with $t\in\left[0,\Delta t_{k}\right)$,
then $t$ denotes the time lag between appearing time and $t_{k}$.
Density (\ref{eq:single_target_density_given_t}) represents the single-target
density at time step $t_{k}$ given that the target appeared with
a time lag $t$. 

\subsection{CD-TPMBM filter\label{subsec:CD-TPMBM-filter}}

As the discretised dynamic model in Section \ref{subsec:Continuous-discrete-model}
is a standard multi-target dynamic model, the posterior and predicted
densities on the set of all trajectories (which include alive and
dead trajectories) are PMBMs \cite{Granstrom18,Angel20_e}. 

Given $\mathbf{z}_{1:k'}$ with $k'\in\left\{ k-1,k\right\} $, the
density $f_{k|k'}\left(\cdot\right)$ of the set $\mathbf{X}_{k}$
of all trajectories up to the current time step $k$ is a PMBM \cite{Granstrom18,Xia19_b,Granstrom19_prov2,Angel20_e}
\begin{align}
f_{k|k'}\left(\mathbf{X}_{k}\right) & =\sum_{\mathbf{X}^{\mathrm{u}}\uplus\mathbf{X}^{\mathrm{d}}=\mathbf{X}_{k}}f_{k|k'}^{\mathrm{p}}\left(\mathbf{X}^{\mathrm{u}}\right)f_{k|k'}^{\mathrm{mbm}}\left(\mathbf{X}^{\mathrm{d}}\right)\label{eq:TPMBM_density}\\
f_{k|k'}^{\mathrm{p}}\left(\mathbf{X}^{\mathrm{u}}\right) & =e^{-\int\lambda_{k|k'}\left(X\right)dX}\prod_{X\in\mathbf{\mathbf{X}^{\mathrm{u}}}}\lambda_{k|k'}\left(X\right)\\
f_{k|k'}^{\mathrm{mbm}}\left(\mathbf{X}^{\mathrm{d}}\right) & =\sum_{a\in\mathcal{A}_{k|k'}}w_{k|k'}^{a}\sum_{\uplus_{l=1}^{n_{k|k'}}\mathbf{X}^{l}=\mathbf{X}^{\mathrm{d}}}\prod_{i=1}^{n_{k|k'}}f_{k|k'}^{i,a^{i}}\left(\mathbf{X}^{i}\right)
\end{align}
where in (\ref{eq:TPMBM_density}) we sum over all disjoint and possibly
empty sets $\mathbf{X}^{\mathrm{u}}$ and $\mathbf{X}^{\mathrm{d}}$
such that $\mathbf{X}^{\mathrm{u}}\cup\mathbf{X}^{\mathrm{d}}=\mathbf{X}_{k}$,
and
\begin{align}
f_{k|k'}^{i,a^{i}}\left(\mathbf{X}\right) & =\begin{cases}
1-r_{k|k'}^{i,a^{i}} & \mathbf{X}=\emptyset\\
r_{k|k'}^{i,a^{i}}p_{k|k'}^{i,a^{i}}\left(X\right) & \mathbf{X}=\left\{ X\right\} \\
0 & \mathrm{otherwise}.
\end{cases}\label{eq:Bernoulli_density_filter}
\end{align}

We proceed to describe the aspects of (\ref{eq:TPMBM_density}) that
are relevant to this work. Details can be found in \cite{Granstrom18}.
The density $f_{k|k'}\left(\cdot\right)$ is the union of two independent
random finite sets: a PPP with density $f_{k|k'}^{\mathrm{p}}\left(\cdot\right)$
and intensity $\lambda_{k|k'}\left(\cdot\right)$, and a multi-Bernoulli
mixture (MBM) with density $f_{k|k'}^{\mathrm{mbm}}\left(\cdot\right)$.
The PPP contains information on trajectories that have never been
detected, but have been discretised at in-sequence measurements, see
Figure \ref{fig:Illustration-trajectories}. The number of potential
trajectories that have ever been present and detected in the surveillance
area is $n_{k|k'}$, which is the number of Bernoullis in each MBM
component. Each received measurement generates one of these potential
trajectories, which are indexed by $i$. A global hypothesis is $a=\left(a^{1},...,a^{n_{k|k'}}\right)$,
where $a^{i}\in\left\{ 1,...,h^{i}\right\} $ is the index to the
local hypothesis for the $i$-th potential trajectory and $h^{i}$
is the number of local hypotheses. Each global hypothesis corresponds
to a multi-Bernoulli in the MBM, and indicates a possible way to associate
the received measurements so far to potential trajectories. The density
of the $i$-th potential trajectory with local hypothesis $a^{i}$
is Bernoulli $f_{k|k'}^{i,a^{i}}\left(\cdot\right)$, whose probability
of existence is $r_{k|k'}^{i,a^{i}}$ and its single-trajectory density
is $p_{k|k'}^{i,a^{i}}\left(\cdot\right)$. The set of all global
hypotheses is $\mathcal{A}_{k|k'}$ \cite{Williams15b}.

The TPMBM posterior (\ref{eq:TPMBM_density}) can be calculated recursively
via a prediction and an update step \cite{Granstrom18,Xia19_b}. The
prediction step is performed as in the TPMBM filter using the corresponding
(interval dependent) probability of survival, single-target transition
density and intensity of new born targets, see (\ref{eq:Probability_survival}),
(\ref{eq:single_target_transition_density}) and (\ref{eq:intensity_birth}).
The update step is similar to the TPMBM filter update.

\section{CD-TPMBM update with OOS measurements\label{sec:TPMBM-update-OoS}}

This section explains the Bayesian processing of OOS measurements
based on the posterior (\ref{eq:TPMBM_density}). Section \ref{subsec:Retrodicted-set-trajectories}
defines the retrodicted set of trajectories. Section \ref{subsec:Retrodiction-step}
and \ref{subsec:Update-step} explain the retrodiction and update
steps. Section \ref{subsec:Marginalisation} addresses the marginalisation
step.

\subsection{Retrodicted set of trajectories\label{subsec:Retrodicted-set-trajectories}}

We consider we know the PMBM posterior over the set $\mathbf{X}_{k}$
of all (sampled) trajectories up to the current time $t_{k}$, $f_{k|k}\left(\cdot\right)$
in (\ref{eq:TPMBM_density}). We receive an OOS set of measurements
with time stamp $\tau$, such that $t_{0}<\tau<t_{k}$. The closest
previously sampled time steps to $\tau$ are $k^{o}-1$ and $k^{o}$,
with continuous times $t_{k^{o}-1}<\tau$ and $t_{k^{o}}>\tau$. We
denote $\Delta t_{o,1}=\tau-t_{k^{o}-1}$ and $\Delta t_{o,2}=t_{k^{o}}-\tau$. 

To perform the update with this OOS set of measurements, we first
perform a retrodiction step in which we calculate the density of the
retrodicted set $\mathbf{Y}_{k}$ of trajectories, e.g., the set of
all trajectories including trajectory state information at time $\tau$.
The set $\mathbf{Y}_{k}$ can be written as $\mathbf{Y}_{k}=\mathbf{X}_{k}^{\mathrm{a}}\cup\mathbf{N}$,
where $\mathbf{X}_{k}^{\mathrm{a}}$ corresponds to the set $\mathbf{X}_{k}$
with additional state information at time $\tau$, and $\mathbf{N}$
denotes the set of trajectories that existed at time step $\tau$,
and appeared and disappeared between time steps $k^{o}-1$ and $k^{o}$.
The trajectories in $\mathbf{N}$ do not belong to $\mathbf{X}_{k}$,
see Figure \ref{fig:Illustration-trajectories}, and we refer to them
as OOS new trajectories at time $\tau$. 

We denote the retrodicted trajectories as $\left(u,Y\right)\in\mathbf{Y}_{k}$,
where mark $u=0$ if the trajectory $Y=\left(\beta,x^{1:\nu}\right)$
does not exist at time $\tau$ (but exists at other sampled times)
and $u=1$ if $Y=\left(\beta,x^{1:\nu}\right)$ exists at time $\tau$,
being the last state $x^{\nu}$ its state at time $\tau$. More information
on marks and point processes can be found at \cite[Chap.8]{Streit_book10}. 

For notational convenience, we write $\left(u,\left(\beta,x^{1:\nu}\right)\right)=\left(u,\beta,x^{1:\nu}\right)$.
In particular, if the trajectory $\left(\beta,x^{1:\nu}\right)\in\mathbf{X}_{k}$
does not exist at time $\tau$, it is included in $\mathbf{X}_{k}^{\mathrm{a}}$
as $\left(0,\beta,x^{1:\nu}\right)$. If the trajectory $\left(\beta,x^{1:\nu}\right)\in\mathbf{X}_{k}$
exists at time $\tau$, it is included in $\mathbf{X}_{k}^{\mathrm{a}}$
as $\left(1,\beta,x^{1:\nu},y\right)$, where its last state $y$
is the state at time $\tau$. These two possibilities are modelled
by a transition density $g_{\tau,k|k}\left(\cdot|X\right)$ that converts
each trajectory $X\in\mathbf{X}_{k}$ into $\left(u,Y\right)\in\mathbf{X}_{k}^{\mathrm{a}}$.
As we explain in Section \ref{subsec:Retrodiction-step}, the set
$\mathbf{N}$ is a PPP independent of $\mathbf{X}_{k}^{\mathrm{a}}$
and a trajectory $\left(u,Y\right)\in\mathbf{N}$ is represented as
$\left(1,\beta,x\right)$, where we set $u=1$ and $\beta=-1$ to
mark that it is an OOS new trajectory. We proceed to illustrate with
an example how sets $\mathbf{X}_{k}^{\mathrm{a}}$ and $\mathbf{N}$
are formed.
\begin{example}
Let us consider we have the trajectories in Example \ref{exa:Set_trajectories}
and Figure \ref{fig:Illustration-trajectories}. We receive an OOS
measurement at time $\tau=6\,\mathrm{s}$. The trajectory that appeared
first $\left(1,\left(1.16,1.34\right)\right)\in\mathbf{X}_{k}$ does
not exist at $\tau$ so it in included in $\mathbf{X}_{k}^{\mathrm{a}}$
as $\left(0,1,\left(1.16,1.34\right)\right)$. The trajectory on top
in Figure \ref{fig:Illustration-trajectories}, $\left(2,\left(2.55,2.63,2.63,2.76,3.05\right)\right)\in\mathbf{X}_{k}$
exists at $\tau$ so it is included in $\mathbf{X}_{k}^{a}$ as $\left(1,2,\left(2.55,2.63,2.63,2.76,3.05,2,96\right)\right)$,
where $2.96$ is the trajectory state at $\tau$. The blue trajectory
was not previously sampled and exists at $\tau$, so it belongs to
$\mathbf{N}$ and has a state $\left(1,-1,2.1\right)$. $\diamondsuit$
\end{example}
The single retrodicted trajectory space is then
\[
\uplus_{u=0}^{1}\uplus_{\left(\beta,\nu\right)\in I_{(k,u)}}\left\{ u\right\} \times\left\{ \beta\right\} \times\mathbb{R}^{\nu n_{x}},
\]
where $I_{(k,0)}=I_{(k)}$ and $I_{(k,1)}=\left\{ \left(-1,1\right)\right\} \cup\left\{ \left(\beta,\nu\right):0\leq\beta\leq k\,\mathrm{and}\,1\leq\nu\leq k-\beta+2\right\} $.
For a real-valued function $\pi\left(\cdot\right)$ on the single-retrodicted
space, its integral is
\begin{align}
\int\pi\left(u,Y\right)d\left(u,Y\right) & =\sum_{u=0}^{1}\sum_{\left(\beta,\nu\right)\in I_{(k,u)}}\int\pi\left(u,\beta,x^{1:\nu}\right)dx^{1:\nu}.\label{eq:single_trajectory_integral_with_u}
\end{align}

\subsection{Retrodiction step\label{subsec:Retrodiction-step}}

Given a trajectory $Y=\left(\beta,x^{1:\nu}\right)$, $\nu>1$, the
trajectory without the last state is denoted by $Y^{-}=\left(\beta,x^{1:\nu-1}\right)$.
We also use symbols $\wedge$ and $\vee$ to denote ``and'' and
``or'', respectively. The transition density to obtain $\left(u,Y\right)\in\mathbf{X}_{k}^{\mathrm{a}}$
from $X\in\mathbf{X}_{k}$ is provided in the following proposition.
\begin{prop}
\label{prop:Transition_density}The transition density $g_{\tau,k|k}\left(\cdot|X\right)$
to augment each trajectory $X=\left(\beta,x^{1:\nu}\right)\in\mathbf{X}_{k}$
with state information at OOS time $\tau$ and produce $\left(u,Y\right)\in\mathbf{X}_{k}^{\mathrm{a}}$
is 
\begin{align}
 & g_{\tau,k|k}\left(u,Y|X\right)\nonumber \\
 & =\begin{cases}
\delta_{X}\left(Y\right)\delta_{0}\left[u\right] & \beta>k^{o}\,\vee\,\omega<k^{o}-1\\
\delta_{X}\left(Y^{-}\right)\\
\times p\left(y|x^{k^{o}-\beta},x^{k^{o}-\beta+1}\right)\delta_{1}\left[u\right] & \beta\leq k^{o}-1\,\wedge\,\omega\geq k^{o}\\
\left(1-p_{1}^{S,o}\right)\delta_{X}\left(Y\right)\delta_{0}\left[u\right]+p_{1}^{S,o}\\
\times\delta_{X}\left(Y^{-}\right)g_{\left(\Delta t_{o,1}\right)}\left(y|x^{\nu}\right)\delta_{1}\left[u\right] & \omega=k^{o}-1\\
\left(1-p_{2}^{S,o}\right)\delta_{X}\left(Y\right)\delta_{0}\left[u\right]\\
+p_{2}^{S,o}\delta_{X}\left(Y^{-}\right)p\left(y|x^{1}\right)\delta_{1}\left[u\right] & \beta=k^{o}
\end{cases}\label{eq:transition_OOS}
\end{align}
where $y$ is the last state of $Y$, $\omega=\beta+\nu-1$, $p_{1}^{S,o}=p_{k}^{S,o}\left(\Delta t_{o,1}\right)$,
$p_{2}^{S,o}=p_{k}^{S,o}\left(\Delta t_{o,2}\right)$ with
\begin{align}
p_{k}^{S,o}\left(\Delta t\right) & =\frac{e^{-\mu\Delta t}-e^{-\mu\Delta t_{k^{o}}}}{1-e^{-\mu\Delta t_{k^{o}}}}\label{eq:p_s_OOS}
\end{align}
and 
\begin{align}
 & p\left(y|x^{k^{o}-\beta},x^{k^{o}-\beta+1}\right)\nonumber \\
 & =\frac{g_{\left(\Delta t_{o,2}\right)}\left(x^{k^{o}-\beta+1}|y\right)g_{\left(\Delta t_{o,1}\right)}\left(y|x^{k^{o}-\beta}\right)}{g_{\left(\Delta t_{k^{o}}\right)}\left(x^{k^{o}-\beta+1}|x^{k^{o}-\beta}\right)}\label{eq:transition_OOS_alive_alive}
\end{align}
\begin{align}
p\left(y|x^{1}\right) & =\frac{g_{\left(\Delta t_{o,2}\right)}\left(x^{1}|y\right)\int_{0}^{\Delta t_{o,1}}p\left(y\left|t\right.\right)p_{\left(\Delta t_{o,1}\right)}\left(t\right)dt}{\int g_{\left(\Delta t_{o,2}\right)}\left(x^{1}|y\right)\int_{0}^{\Delta t_{o,1}}p\left(y\left|t\right.\right)p_{\left(\Delta t_{o,1}\right)}\left(t\right)dtdy}.\label{eq:transition_OOS_dead_alive}
\end{align}
\end{prop}
The first entry in (\ref{eq:transition_OOS}) indicates that if a
trajectory $X$ was born after $k^{o}$ or its final time step $\omega$
occurred before $k^{o}-1$, its state does not exist at time  $\tau$
with with probability one. That is, this entry considers trajectories
that do not exist at time $\tau$ but exist at other sampled time
steps. The second entry in (\ref{eq:transition_OOS}) indicates that
if a trajectory $X$ was born at time step $k^{o}-1$, or earlier,
and finished at time step $k^{o}$, or afterwards, then the trajectory
exists at time $\tau$ with probability one. In addition, given the
states of $X$ at time steps $k^{o}-1$ and $k^{o}$, its state at
at time $\tau$ can be directly obtained using Bayes' rule and the
properties of the discretised single-target transition density (\ref{eq:single_target_transition_density}),
resulting in (\ref{eq:transition_OOS_alive_alive}). The third entry
in (\ref{eq:transition_OOS}) considers a trajectory $X$ that finished
at time step $k^{o}-1$. This trajectory disappeared (in continuous
time) at any time between $t_{k^{o}-1}$ and $t_{k^{o}}$, and the
probability that it disappeared between times $\tau$ and $t_{k^{o}}$,
which implies that it existed at time $\tau$, is $p_{1}^{S,o}$.
If it exists, its state is obtained using the single-target transition
density (\ref{eq:single_target_transition_density}) with a time interval
$\Delta t_{o,1}$. The fourth entry in (\ref{eq:transition_OOS})
considers a trajectory $X$ that was born at time step $k^{o}$. This
trajectory appeared (in continuous time) at any time between $t_{k^{o}-1}$
and $t_{k^{o}}$, and the probability that it appeared between times
$t_{k^{o}-1}$ and $\tau$, which implies that it existed at time
$\tau$, is $p_{2}^{S,o}$. If it exist, its state at $\tau$ is given
by applying Bayes' rule to its prior density at time step $\tau$
corrected by the information provided by its state at time $t_{k^{o}}$.
The resulting transition density is (\ref{eq:transition_OOS_dead_alive}).
More details on how to calculate $p_{1}^{S,o}$ and $p_{2}^{S,o}$
are provided in Appendix \ref{sec:Appendix_A}.

Once we have the transition density for the retrodiction step, we
can obtain the PMBM retrodiction step via the following theorem.
\begin{thm}
\label{thm:Retrodiction-PMBM}Given the PMBM posterior $f_{k|k}\left(\cdot\right)$
in (\ref{eq:TPMBM_density}) on the set of all sampled trajectories,
the retrodicted density on the set $\mathbf{Y}_{k}$ of trajectories
augmented with information at time $\tau<t_{k}$ is a PMBM with density
\begin{align}
f_{\tau,k|k}\left(\mathbf{Y}_{k}\right) & =\sum_{\mathbf{Y}^{\mathrm{u}}\uplus\mathbf{Y}^{\mathrm{d}}=\mathbf{Y}_{k}}f_{\tau,k|k}^{\mathrm{p}}\left(\mathbf{Y}^{\mathrm{u}}\right)f_{\tau,k|k}^{\mathrm{mbm}}\left(\mathbf{Y}^{\mathrm{d}}\right)\label{eq:TPMBM_retrodiction_at_OOS}\\
f_{\tau,k|k}^{\mathrm{p}}\left(\mathbf{Y}^{\mathrm{u}}\right) & =e^{-\int\lambda_{\tau,k|k}\left(u,Y\right)d\left(u,Y\right)}\prod_{\left(u,Y\right)\in\mathbf{Y}^{\mathrm{u}}}\lambda_{\tau,k|k}\left(u,Y\right)\\
f_{\tau,k|k}^{\mathrm{mbm}}\left(\mathbf{Y}^{\mathrm{d}}\right) & =\sum_{a\in\mathcal{A}_{k|k}}w_{k|k}^{a}\sum_{\uplus_{l=1}^{n_{k|k}}\mathbf{Y}^{l}=\mathbf{Y}^{\mathrm{d}}}\prod_{i=1}^{n_{k|k}}f_{\tau,k|k}^{i,a^{i}}\left(\mathbf{Y}^{i}\right)
\end{align}
where the intensity of the PPP \textup{$f_{\tau,k|k}^{\mathrm{p}}\left(\cdot\right)$}
is
\begin{align}
\lambda_{\tau,k|k}\left(u,Y\right) & =\lambda_{\tau,k|k}^{B}\left(u,Y\right)\nonumber \\
 & \;+\int g_{\tau,k|k}\left(u,Y|X\right)\lambda_{k|k}\left(X\right)dX,\label{eq:intensity_OOS}
\end{align}
and the intensity of OOS new trajectories is
\begin{align}
\lambda_{\tau,k|k}^{B}\left(u,\beta,x^{1:\nu}\right) & =w^{B}\left(\Delta t_{o,1},\Delta t_{o,2}\right)\delta_{1}\left[u\right]\delta_{-1}\left[\beta\right]\nonumber \\
 & \times\delta_{1}\left[\nu\right]\int_{0}^{\Delta t_{o,1}}p\left(x^{1}\left|t\right.\right)p_{\left(\Delta t_{o,1}\right)}\left(t\right)dt.\label{eq:intensity_birth_OOS}\\
w^{B}\left(\Delta t_{o,1},\Delta t_{o,2}\right) & =\frac{\lambda}{\mu}\left(1-e^{-\mu\Delta t_{o,1}}\right)\left(1-e^{-\mu\Delta t_{o,2}}\right).\label{eq:mean_number_OOS_trajectories}
\end{align}
The probability of existence and single-target density of Bernoulli
$f_{\tau,k|k}^{i,a^{i}}\left(\cdot\right)$ are
\begin{align}
r_{\tau,k|k}^{i,a^{i}} & =r_{k|k}^{i,a^{i}}\label{eq:existence_OOS}\\
p_{\tau,k|k}^{i,a^{i}}\left(u,Y\right) & =\int g_{\tau,k|k}\left(u,Y|X\right)p_{k|k}^{i,a^{i}}\left(X\right)dX.\label{eq:single_target_Bernoulli_OOS}
\end{align}
\end{thm}
Theorem \ref{thm:Retrodiction-PMBM} is proved in Appendix \ref{sec:Appendix_A},
and results from the application of the single-trajectory transition
density in (\ref{prop:Transition_density}) to a PMBM density (\ref{eq:TPMBM_density}),
accounting for the distribution of the set $\mathbf{N}$ of OOS new
trajectories, which is a PPP with intensity $\lambda_{\tau,k|k}^{B}\left(\cdot\right)$. 

The probability of existence of the Bernoulli components does not
change, see (\ref{eq:existence_OOS}). The reason is that all the
trajectories that belong to $\mathbf{X}_{k}$ also belong to $\mathbf{Y}_{k}$,
so there is no change in their probability of existence. A similar
phenomenon happens in the TPMBM prediction step when we consider all
trajectories \cite{Granstrom18,Angel20_e}. The single-target densities
(\ref{eq:single_target_Bernoulli_OOS}) are transformed using the
transition density $g_{\tau,k|k}\left(\cdot|\cdot\right)$, which
augments trajectories with state information at time $\tau$. The
intensity of the PPP (\ref{eq:intensity_OOS}) is the sum of the intensity
$\lambda_{\tau,k|k}^{B}\left(\cdot\right)$ and the intensity of the
undetected trajectories in $\mathbf{X}_{k}$ augmented with information
at time $\tau$. Equation (\ref{eq:mean_number_OOS_trajectories})
represents the expected number of OOS new trajectories. This number
is the expected number of trajectories that appear in an interval
$\Delta t_{o,1}$ and are alive at its end, which is given by $\frac{\lambda}{\mu}\left(1-e^{-\mu\Delta t_{o,1}}\right)$
\cite{Angel20,Kulkarni_book16}, multiplied by the probability that
a trajectory disappears in an interval $\Delta t_{o,2}$, which is
given by $\left(1-e^{-\mu\Delta t_{o,2}}\right)$, see (\ref{eq:Probability_survival}). 

\begin{figure}
\begin{centering}
\includegraphics[scale=0.6]{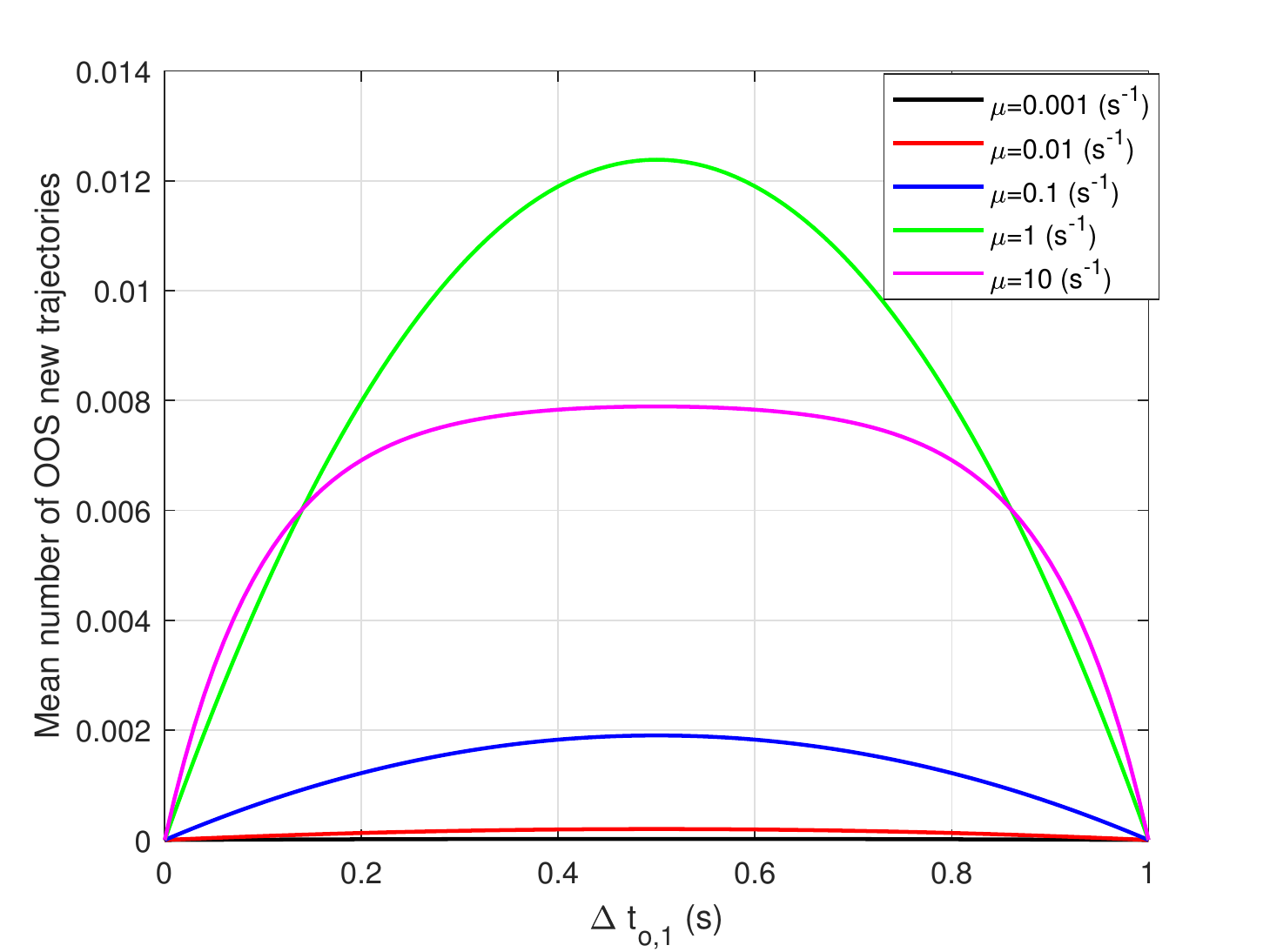}
\par\end{centering}
\caption{\label{fig:Mean-number-OOS-trajectories}Mean number of OOS new trajectories,
see (\ref{eq:mean_number_OOS_trajectories}), as a function of $\Delta t_{o,1}$
for a time interval $t_{k^{o}}-t_{k^{o}-1}=1\,\mathrm{s}$ and $\lambda=0.08\,\mathrm{s}^{-1}$.
The maximum is achieved for the middle of the time interval $\Delta t_{o,1}=0.5\,\mathrm{s}$.
}
\end{figure}

We plot the mean number of OOS new trajectories, see (\ref{eq:mean_number_OOS_trajectories}),
as a function of $\Delta t_{o,1}$ in one illustrative example in
Figure \ref{fig:Mean-number-OOS-trajectories}. The maximum is obtained
at the middle of the interval $\Delta t_{o,1}=\Delta t_{k^{o}}/2$,
which can also be proved analytically. This means that if the OOS
measurement falls in the middle of two sampled times, the number of
OOS new trajectories is at its maximum. The mean number of OOS new
trajectories increases with $\lambda$, as more targets appear in
the scene. In addition, the mean number of OOS new trajectories initially
increases with $\mu$, but it then decreases. We recall that $1/\mu$
is the expected life span of the trajectories \cite[Sec. II]{Angel20}.
For sufficiently small $\mu$, targets that appeared between $t_{k^{o}-1}$
and $\tau$ are still alive at $t_{k^{o}}$ with high probability
and $w^{B}\left(\cdot\right)$ is small. As $\mu$ starts increasing,
the probability that these targets are not alive at $t_{k^{o}}$ increases,
and therefore, $w^{B}\left(\cdot\right)$ increases. However, as $\mu$
increases the number of targets that appear between $t_{k^{o}-1}$
and $\tau$ and are alive at $\tau$ also decreases, which implies
that $w^{B}\left(\cdot\right)$ starts to decrease after a certain
point. 

\subsection{Update step\label{subsec:Update-step}}

The measurement model at time $\tau$, see Section \ref{subsec:Sets-of-targets},
can be written in terms of $\mathbf{Y}_{k}$, as follows. Each trajectory
$\left(u,\beta,x^{1:\nu}\right)\in\mathbf{Y}_{k}$ is detected with
probability
\begin{align}
p^{D}\left(u,\beta,x^{1:\nu}\right) & =\begin{cases}
p^{D}\left(x^{\nu}\right) & u=1\\
0 & \mathrm{otherwise}
\end{cases}
\end{align}
and generates a measurement with density $l\left(\cdot|u,\beta,x^{1:\nu}\right)=l\left(\cdot|x^{\nu}\right)$,
or misdetected with probability $1-p^{D}\left(u,\beta,x^{1:\nu}\right)$.
The clutter model remains unchanged.

For this measurement model and a PMBM prior (\ref{eq:TPMBM_retrodiction_at_OOS}),
the updated density $f_{\tau,k|,\tau,k}\left(\cdot\right)$ is also
PMBM \cite{Williams15b,Granstrom18,Angel20_e}. The update is analogous
to the trajectory PMBM filter update \cite{Granstrom18,Angel20_e},
but using the retrodicted trajectory integral (\ref{eq:single_trajectory_integral_with_u}).

\subsection{Marginalisation\label{subsec:Marginalisation}}

The steps in Sections \ref{subsec:Retrodiction-step} and \ref{subsec:Update-step}
provide the closed-form update when we receive the first OOS set of
measurements. In order to continue with the filtering recursion, we
can proceed in two forms. One is to transform the augmented trajectories
$\left(u,Y\right)$ into trajectories of the type $\left(\beta,x^{1:\nu}\right)$,
with the states arranged in consecutive time steps. In order to do
this, the time index $k$ of the measurements changes as we insert
a new measurement in the previous sequence. In addition, the meaning
of $\beta$ changes. For trajectories born the time step corresponding
to $\tau$, it represents trajectories in $\mathbf{N}$, which appear
and disappear in between the two closest sampling times. If we follow
this approach, it is possible to generalise the process in this section
to deal with OOS measurements in an exact way. 

However, as most of the measurements are expected to be in-sequence,
we pursue the simpler approach of marginalising out the information
at time $\tau$. That is, once we have updated all trajectory information
based on the OOS measurement, we only keep the information at the
in-sequence measurement sampled times, not at OOS measurement times.
This marginalisation can be obtained by applying the transition density

\begin{align}
g_{m}\left(\mathbf{X}|\left(u,\beta,x^{1:\nu}\right)\right) & =\begin{cases}
\delta_{\emptyset}\left(\mathbf{X}\right) & \beta=-1\\
\delta_{\left\{ \left(\beta,x^{1:\nu}\right)\right\} }\left(\mathbf{X}\right) & \beta>-1,u=0\\
\delta_{\left\{ \left(\beta,x^{1:\nu-1}\right)\right\} }\left(\mathbf{X}\right) & \beta>-1,u=1\\
0 & \mathrm{otherwise}
\end{cases}
\end{align}
to each trajectory in $\mathbf{Y}_{k}$. This transition density
is actually a Bernoulli transition density with state dependent probability
of survival. When applied to the updated PMBM, the result is a PMBM
that discards trajectory information at time $\tau$ \cite{Granstrom19_prov2,Granstrom20b}. 

Every time we receive an OOS measurement, we perform the steps of
retrodiction, update and marginalisation. The procedure provides the
exact solution posterior at the in-sequence sampling times unless
we get more than one OOS set of measurements in the same time interval
$\left(t_{k^{o}-1},t_{k^{o}}\right)$. In this case, the procedure
is an approximation as, for the exact retrodiction in Theorem \ref{thm:Retrodiction-PMBM},
we require access to the two closest states of the trajectories at
the time steps when we have received measurements.

\section{OOS measurement processing with Gaussian CD-TPMBM implementation\label{sec:OOS-Gaussian}}

This section explains how to process OOS measurements with a Gaussian
implementation of the CD-TPMBM filter. The single-target models are
explained in Section \ref{subsec:Single-target-models}, the Gaussian
TPMBM posterior in Section \ref{subsec:Gaussian-TPMBM-posterior},
the retrodiction step in Section \ref{subsec:OOS-Gaussian-retrodiction-step}.
Finally, practical aspects are discussed in Section \ref{subsec:Practical-considerations}.

\subsection{Single-target models\label{subsec:Single-target-models}}

For the Gaussian implementation, we use a linear/Gaussian measurement
model $l\left(\cdot|x\right)=\mathcal{N}\left(\cdot;Hx,R\right)$
and a constant probability $p_{D}$ of detection. We consider the
Wiener velocity model \cite{Sarkka_book19} for single-target dynamics
with a single-target state 
\begin{align}
x\left(t_{k}\right) & =\left[p_{1}\left(t_{k}\right),...,p_{d}\left(t_{k}\right),v_{1}\left(t_{k}\right),...,v_{d}\left(t_{k}\right)\right]^{T}
\end{align}
where $d=n_{x}/2$ is the dimension of the space where the target
moves. For this dynamic model, we can obtain a best Gaussian fit to
the PPP of new born targets that enables Gaussian implementations
\cite[Prop. 2]{Angel20}. This result directly extends to the PPP
of OOS new trajectories in (\ref{eq:intensity_birth_OOS}). 

For the Wiener velocity model, we also have \cite{Sarkka_book19}
\begin{align}
F_{\left(\Delta t_{k}\right)} & =\left(\begin{array}{cc}
I_{d} & \Delta t_{k}I_{d}\\
0_{d} & I_{d}
\end{array}\right)\label{eq:F_t_Wiener_velocity}\\
Q_{\left(\Delta t_{k}\right)} & =q\left(\begin{array}{cc}
\frac{\left(\Delta t_{k}\right)^{3}}{3}I_{d} & \frac{\left(\Delta t_{k}\right)^{2}}{2}I_{d}\\
\frac{\left(\Delta t_{k}\right)^{2}}{2}I_{d} & \Delta t_{k}I_{d}
\end{array}\right)\label{eq:Q_t_Wiener_velocity}
\end{align}
where $q$ is a model parameter.

\subsection{Gaussian TPMBM posterior\label{subsec:Gaussian-TPMBM-posterior}}

For the models explained in Section \ref{subsec:Single-target-models},
we can use the Gaussian implementation of the TPMBM filter for the
set of all trajectories in \cite{Angel20_e}. We proceed to describe
the main aspects. Details can be found in \cite{Angel20_e,Granstrom18}. 

A Gaussian density in the single-trajectory space is
\begin{align}
\mathcal{N}\left(\beta,x^{1:\nu};\overline{\beta},\overline{x},P\right) & =\begin{cases}
\mathcal{N}\left(x^{1:\nu};\overline{x},P\right) & \beta=\overline{\beta},\,\nu=\iota\\
0 & \mathrm{otherwise}
\end{cases}\label{eq:Trajectory_Gaussian}
\end{align}
where $\iota=\mathrm{dim}\left(\overline{x}\right)/n_{x}$ and $\mathrm{dim}\left(\overline{x}\right)$
is the dimension of $\overline{x}$. Equation (\ref{eq:Trajectory_Gaussian})
represents a Gaussian trajectory density with start time $\overline{\beta}$,
duration $\iota$, mean $\overline{x}\in\mathbb{R}^{\iota n_{x}}$
and covariance matrix $P\in\mathbb{R}^{\iota n_{x}\times\iota n_{x}}$
evaluated at $\left(\beta,x^{1:\nu}\right)$.

The $i$-th Bernoulli component with local hypothesis $a^{i}$ has
a single-trajectory density
\begin{align}
p_{k|k}^{i,a^{i}}\left(X\right) & =\sum_{\kappa=\beta^{i,a^{i}}}^{k}\alpha_{k|k}^{i,a^{i}}\left(\kappa\right)\mathcal{N}\left(X;\beta^{i,a^{i}},\overline{x}_{k|k}^{i,a^{i}}\left(\kappa\right),P_{k|k}^{i,a^{i}}\left(\kappa\right)\right)\label{eq:single_trajectory_Gaussian_all}
\end{align}
where $\beta^{i,a^{i}}$ is the start time, $\alpha_{k|k}^{i,a^{i}}\left(\kappa\right)$
is the probability that the corresponding trajectory terminates at
time step $\kappa$ (conditioned on existence), and $\overline{x}_{k|k}^{i,a^{i}}\left(\kappa\right)\in\mathbb{R}^{\iota n_{x}}$
and $P_{k|k}^{i,a^{i}}\left(\kappa\right)\in\mathbb{R}^{\iota n_{x}\times\iota n_{x}}$,
with $\iota=\kappa-\beta^{i,a^{i}}+1$, are the mean and the covariance
matrix of the trajectory given that it ends at time step $\kappa$.
The coefficients $\alpha_{k|k}^{i,a^{i}}\left(\kappa\right)$, $\kappa=\beta^{i,a^{i}},...,k$,
sum to one.

For simplicity, the intensity of the PPP only considers alive trajectories
and has the form
\begin{align}
\lambda_{k|k}\left(X\right) & =\sum_{q=1}^{n_{k|k}^{p}}w_{k|k}^{p,q}\mathcal{N}\left(X;\beta_{k|k}^{p,q},\overline{x}_{k|k}^{p,q},P_{k|k}^{p,q}\right)\label{eq:intensity_Gaussian_alive}
\end{align}
where $n_{k|k}^{p}$ is the number of components, $w_{k|k}^{p,q}$,
$\beta_{k|k}^{p,q}$, $\overline{x}_{k|k}^{p,q}$ and $P_{k|k}^{p,q}$
are the weight, starting time, mean and covariance matrix of the $q$th
component, respectively. As the PPP trajectories are alive, $\beta_{k|k}^{p,q}+\mathrm{dim}\left(\overline{x}_{k|k}^{p,q}\right)/n_{x}-1=k$.

\subsection{OOS retrodiction step\label{subsec:OOS-Gaussian-retrodiction-step}}

To perform the retrodiction step, we need to calculate (\ref{eq:intensity_OOS})
and (\ref{eq:single_target_Bernoulli_OOS}) when the input is (\ref{eq:single_trajectory_Gaussian_all})
and (\ref{eq:intensity_Gaussian_alive}). These results can be directly
established by calculating the integral (\ref{eq:single_target_Bernoulli_OOS})
for a Gaussian input (\ref{eq:Trajectory_Gaussian}). We denote $F_{1}=F_{\left(\Delta t_{o,1}\right)}$,
$F_{2}=F_{\left(\Delta t_{o,2}\right)}$, $Q_{1}=Q_{\left(\Delta t_{o,1}\right)}$
and $Q_{2}=Q_{\left(\Delta t_{o,2}\right)}$. 

We approximate the integral w.r.t. time in (\ref{eq:transition_OOS_dead_alive})
for the Wiener velocity model by its best Gaussian fit via KLD minimisation.
The resulting moments, called $\overline{x}_{b,1}$ and $P_{b,1}$,
are given by Prop. 2 in \cite{Angel20} using $\Delta t_{o,1}$ as
the time interval. The rest of the calculations are closed-form to
yield this lemma. 
\begin{lem}
\label{lem:Gaussian_OOS_retrodiction}Given $p\left(X\right)=\mathcal{N}\left(X;\overline{\beta},\overline{x},P\right)$
and $g_{\tau,k|k}\left(\cdot|X\right)$ in Prop. \ref{prop:Transition_density}
and the best Gaussian fit to the integral in (\ref{eq:transition_OOS_dead_alive}),
with moments $\overline{x}_{b,1}$ and $P_{b,1}$ \cite[Prop. 2]{Angel20},
the density of its augmented trajectory $\left(u,Y\right)$ is
\begin{align}
 & \int g_{\tau,k|k}\left(u,Y|X\right)p\left(X\right)dX\nonumber \\
 & \:=\begin{cases}
p\left(Y\right)\delta_{0}\left[u\right] & \overline{\beta}>k^{o}\,\vee\,\omega<k^{o}-1\\
\mathcal{N}\left(Y;\overline{\beta},\overline{y}_{pp},P_{pp}\right)\delta_{1}\left[u\right] & \overline{\beta}\leq k^{o}-1\,\wedge\,\omega\geq k^{o}\\
\left(1-p_{1}^{S,o}\right)p\left(Y\right)\delta_{0}\left[u\right]\\
\;+p_{1}^{S,o}\mathcal{N}\left(Y;\overline{\beta},\overline{y}_{pn},P_{pn}\right)\delta_{1}\left[u\right] & \omega=k^{o}-1\\
\left(1-p_{2}^{S,o}\right)p\left(Y\right)\delta_{0}\left[u\right]\\
\;+p_{2}^{S,o}\mathcal{N}\left(Y;\overline{\beta},\overline{y}_{np},P_{np}\right)\delta_{1}\left[u\right] & \overline{\beta}=k^{o}
\end{cases}\label{eq:Gaussian_OOS_retrodiction}
\end{align}
where $\omega=\overline{\beta}+\mathrm{dim}\left(\overline{x}\right)/n_{x}-1$
is the final time step. For $p\left(X\right)$ present at $k^{o}-1$
and $k^{o}$, we have
\begin{align*}
\overline{y}_{pp}=\left[\overline{x}^{T},\left(F_{pp}\overline{x}\right)^{T}\right]^{T} & ,\:P_{pp}=\left[\begin{array}{cc}
P & PF_{pp}^{T}\\
F_{pp}P & F_{pp}PF_{pp}^{T}+Q_{pp}
\end{array}\right]
\end{align*}
\begin{align*}
F_{pp} & =\left[0_{n_{x}\times n_{x}\left(k^{o}-\overline{\beta}-1\right)},F_{1}-K_{pp}F_{2}F_{1},K_{pp},0_{n_{x}\times n_{x}\left(\omega-k^{o}\right)}\right]
\end{align*}
\begin{align}
Q_{pp} & =Q_{1}-K_{pp}F_{2}Q_{1}\\
K_{pp} & =Q_{1}F_{2}^{T}\left(F_{2}Q_{1}F_{2}^{T}+Q_{2}\right)^{-1}.
\end{align}
For $p\left(X\right)$ present at $k^{o}-1$ but not at $k^{o}$,
we have
\begin{align*}
\overline{y}_{pn}=\left[\overline{x}^{T},\left(F_{pn}\overline{x}\right)^{T}\right]^{T} & ,\:P_{pn}=\left[\begin{array}{cc}
P & PF_{pn}^{T}\\
F_{pn}P & F_{pn}PF_{pn}^{T}+Q_{1}
\end{array}\right]
\end{align*}
\begin{align*}
F_{pn} & =\left[0_{n_{x}\times n_{x}\left(\omega-\overline{\beta}\right)},F_{1}\right].
\end{align*}
For $p\left(X\right)$ not present at $k^{o}-1$ but present at $k^{o}$,
we have
\begin{align*}
\overline{y}_{np} & =\left[\overline{x}^{T},\left(\left(I-K_{np}F_{2}\right)\overline{x}_{b,1}+F_{np}\overline{x}\right)^{T}\right]^{T}
\end{align*}
\begin{align*}
P_{np} & =\left[\begin{array}{cc}
P & PF_{np}^{T}\\
F_{np}P & F_{np}PF_{np}^{T}+Q_{np}
\end{array}\right]
\end{align*}
\begin{align}
F_{np} & =\left[K_{np},0_{n_{x}\times n_{x}\left(\omega-\overline{\beta}\right)}\right]\\
Q_{np} & =P_{b,1}-K_{np}F_{2}P_{b,1}\\
K_{np} & =P_{b,1}F_{2}^{T}\left(F_{2}P_{b,1}F_{2}^{T}+Q_{2}\right)^{-1}.
\end{align}
\end{lem}
The proof of Lemma \ref{lem:Gaussian_OOS_retrodiction} is given in
Appendix \ref{sec:Appendix_B}. In the lemma, there is one entry per
each of the entries in the transition density in Prop. \ref{prop:Transition_density}.
The first entry deals with trajectories that start after $k^{o}$
or end before than $k^{o}-1$, which imply that there is no OOS state
and the density remains unchanged. The second entry considers trajectories
that are present at $k^{o}-1$ and $k^{o}$ so the trajectory exists
at the OOS time. The third entry correspond to trajectories that are
present at $k^{o}-1$ but not at $k^{o}$, which implies that the
trajectory is extended with probability $p_{1}^{S,o}$. The fourth
entry represents trajectories not present at $k^{o}-1$ but present
at $k^{o}$, in which case the trajectory is extended with probability
$p_{2}^{S,o}$. 

Applying Lemma \ref{lem:Gaussian_OOS_retrodiction} to each Gaussian
component of the PPP (\ref{eq:intensity_Gaussian_alive}) and the
Bernoulli single-trajectory density in (\ref{eq:single_trajectory_Gaussian_all}),
we obtain the retrodicted PMBM density $f_{\tau,k|k}\left(\cdot\right)$,
see (\ref{eq:TPMBM_retrodiction_at_OOS}) The number of components
in the PPP and in (\ref{eq:single_trajectory_Gaussian_all}) may increase
due to the entries that have two terms in (\ref{eq:Gaussian_OOS_retrodiction}).
After computing $f_{\tau,k|k}\left(\cdot\right)$, we apply the TPMBM
update for a Gaussian implementation with all trajectories, explained
in \cite{Granstrom19_prov2,Angel20_e}, with some minor differences
that are explained in Appendix \ref{sec:Appendix_C}. 

The marginalisation step for PMBMs on sets of trajectories is explained
in \cite{Granstrom20b}. In our case, this step marginalises out variable
$u$ and the state information corresponding to time $\tau$ for each
Gaussian. The result is a Gaussian mixture of the form (\ref{eq:single_trajectory_Gaussian_all}). 

\subsection{Implementation aspects\label{subsec:Practical-considerations}}

In this section, we discuss some aspects required for the implementation
of the proposed OOS update. First of all, we implement the Gaussian
CD-TPMBM for all trajectories in a similar manner as the TPMBM in
\cite{Angel20_e}. That is, to deal with the high number of hypotheses,
we use ellipsoidal gating for the data associations, Murty's algorithm
to select global hypotheses with high weights, and pruning to remove
global hypotheses and PPP components with low weights. If $\alpha_{k|k}^{i,a^{i}}\left(k\right)$
in (\ref{eq:single_trajectory_Gaussian_all}) for a Bernoulli is less
than a threshold $\Gamma_{a}$, we set $\alpha_{k|k}^{i,a^{i}}\left(k\right)=0$,
which implies that it is considered dead at time step $k$ and is
not further propagated through filtering.

The CD-TPMBM filter is implemented using an $L$-scan window. That
is, for each single-trajectory density, the states corresponding to
time steps outside the interval from $k-L+1$ to $k$ are approximated
as independent. This implies that the covariance matrices have a block-diagonal
structure \cite[Eq. (73)]{Angel20_e}. Due to this structure, in our
implementation, we only process a set of OOS measurements if it arrives
inside the $L$-scan window, i.e., $k^{o}\geq k-L+2$. Apart from
the $L$-scan implementation, it is also possible to implement the
Gaussian filters in information form \cite{Granstrom18,Granstrom19_prov2}. 

The CD-TPMB filter is analogous to the CD-TPMBM filter but adding
a projection step after each update to keep the TPMB form \cite{Angel20_e}.
Therefore, we can directly apply the proposed OOS update to a CD-TPMB
filter followed by this projection step after the OOS measurement
update.

\section{Simulations\label{sec:Simulations}}

In this section, we compare the CD-TPMBM and CD-TPMB filters, with
and without OOS measurement processing\footnote{Matlab code is available at https://github.com/Agarciafernandez/MTT.}.
The CD-TPMBM and CD-TPMB filters with the optimal OOS processing explained
above are referred to as OOS-TPMBM and OOS-TPMB filters. If the OOS
measurement time stamp is exactly the time stamp of an in-sequence
measurement, we do not have to account for target appearances and
disappearances at OOS time and proceed as in Sections \ref{sec:TPMBM-update-OoS}
and \ref{sec:OOS-Gaussian}. Instead, we can update each single trajectory
density of the TPMBM filter using the approach in \cite{Koch11}.
Therefore, we consider another baseline algorithm, in which for each
OOS measurement, we calculate the nearest in-sequence measurement
time stamp, and apply the single-trajectory update in \cite{Koch11}.
We use the acronyms (N)OOS-TPMBM and (N)OOS-TPMB to refer to these
variants of the filters. The variants of the filters without OOS measurement
processing simply discard OOS measurements.

The filters have been implemented with the parameters: maximum number
of global hypotheses $N_{h}=200$, threshold for pruning global hypotheses
$10^{-4}$, threshold for PPP pruning $\Gamma_{p}=10^{-5}$, $L\in\left\{ 3,5\right\} $
and $\Gamma_{a}=10^{-4}$. The TPMB filters estimate trajectories
whose existence is higher than 0.5 \cite[Sec. V.D]{Angel20_e} and
the TPMBM filters use Estimator 1 in \cite{Angel18_b} with threshold
0.4. The algorithms are implemented in Matlab with the compiled Murty's
algorithm in \cite{Crouse17}. 

We consider a 2-D scenario with the Wiener velocity model and dynamic
parameters: $\lambda=0.12\,\mathrm{s}^{-1}$, $\mu=0.02\,\mathrm{s}^{-1}$,
$q=0.2\,\mathrm{m}^{2}/\mathrm{s}^{3}$, $d=2$. Thus, the average
life span of a target is $1/\mu=50\,\mathrm{s}$ and, in the stationary
regime of the birth/death process, the number of alive targets is
Poisson distributed with parameter $\frac{\lambda}{\mu}=6$. The prior
moments at appearance time are: $\overline{x}_{a}=\left[\overline{p}_{a}^{T},\overline{v}_{a}^{T}\right]^{T}$
with $\overline{p}_{a}=\left[200,200\right]^{T}\,\left(\mathrm{m}\right)$,
$\overline{v}_{a}=\left[3,0\right]^{T}\,\left(\mathrm{m/s}\right)$,
and $P_{a}=\mathrm{diag}\left(\left[P_{a}^{pp},P_{a}^{vv}\right]\right)$
with $P_{a}^{pp}=\mathrm{diag}\left(\left[50^{2},50^{2}\right]\right)\,\left(\mathrm{m}^{2}\right)$
and $P_{a}^{vv}=\mathrm{diag}\left(\left[1,1\right]\right)\,\left(\mathrm{m}^{2}/\mathrm{s}^{2}\right)$. 

\begin{figure}
\begin{centering}
\includegraphics[scale=0.6]{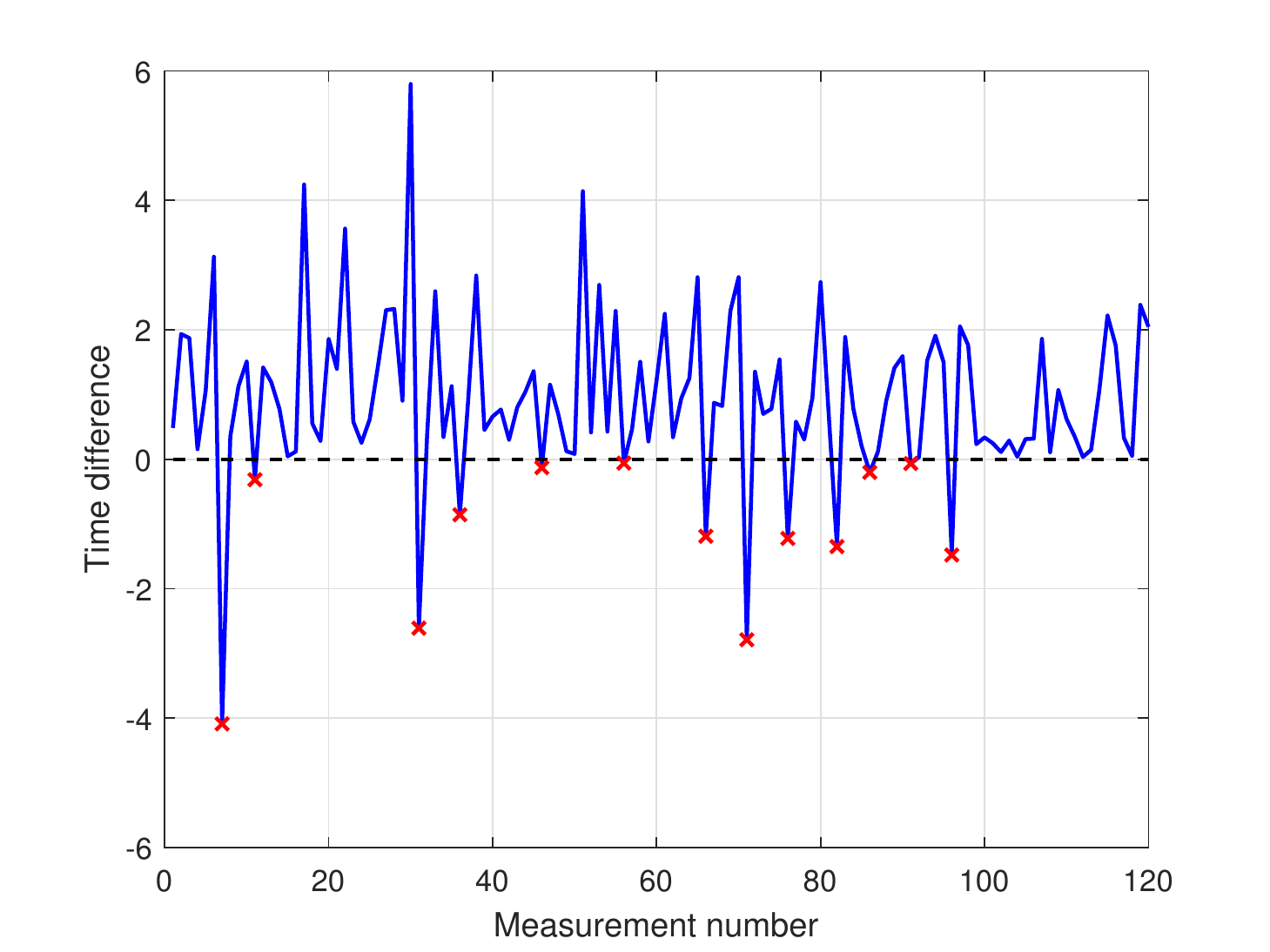}
\par\end{centering}
\caption{\label{fig:Time-difference-measurements}Time difference between received
measurements. OOS measurements have a negative time difference and
are highlighted with a red cross.}

\end{figure}

The sensor measures position with likelihood $l\left(\cdot|x\right)=\mathcal{N}\left(\cdot;Hx,R\right)$,
\begin{align*}
H=\left(\begin{array}{cccc}
1 & 0 & 0 & 0\\
0 & 1 & 0 & 0
\end{array}\right),\quad R=\sigma^{2}I_{2},
\end{align*}
where $\sigma^{2}=4\,\left(\mathrm{m}^{2}\right)$, and $p_{D}=0.9$.
The clutter intensity is $\lambda^{C}\left(z\right)=\overline{\lambda}^{C}u_{A}\left(z\right)$
where $u_{A}\left(\cdot\right)$ is a uniform density in $A=\left[0,800\right]\times\left[0,400\right]\,\left(\mathrm{m}\right)$
and $\overline{\lambda}^{C}=10$. The sensor takes 120 measurements
with a time interval between measurements that is drawn from an exponential
distribution with parameter $\mu_{m}=1\,\mathrm{s}^{-1}$. To simulate
OOS measurements, for every 5 of the 120 measurements, we draw a random
number $n_{o}$ from a Poisson distribution with parameter 1 and place
this measurement $n_{o}$ time steps afterwards. The resulting time
difference between received measurements in our simulation is shown
in Figure \ref{fig:Time-difference-measurements}.

\begin{figure}
\begin{centering}
\includegraphics[scale=0.27]{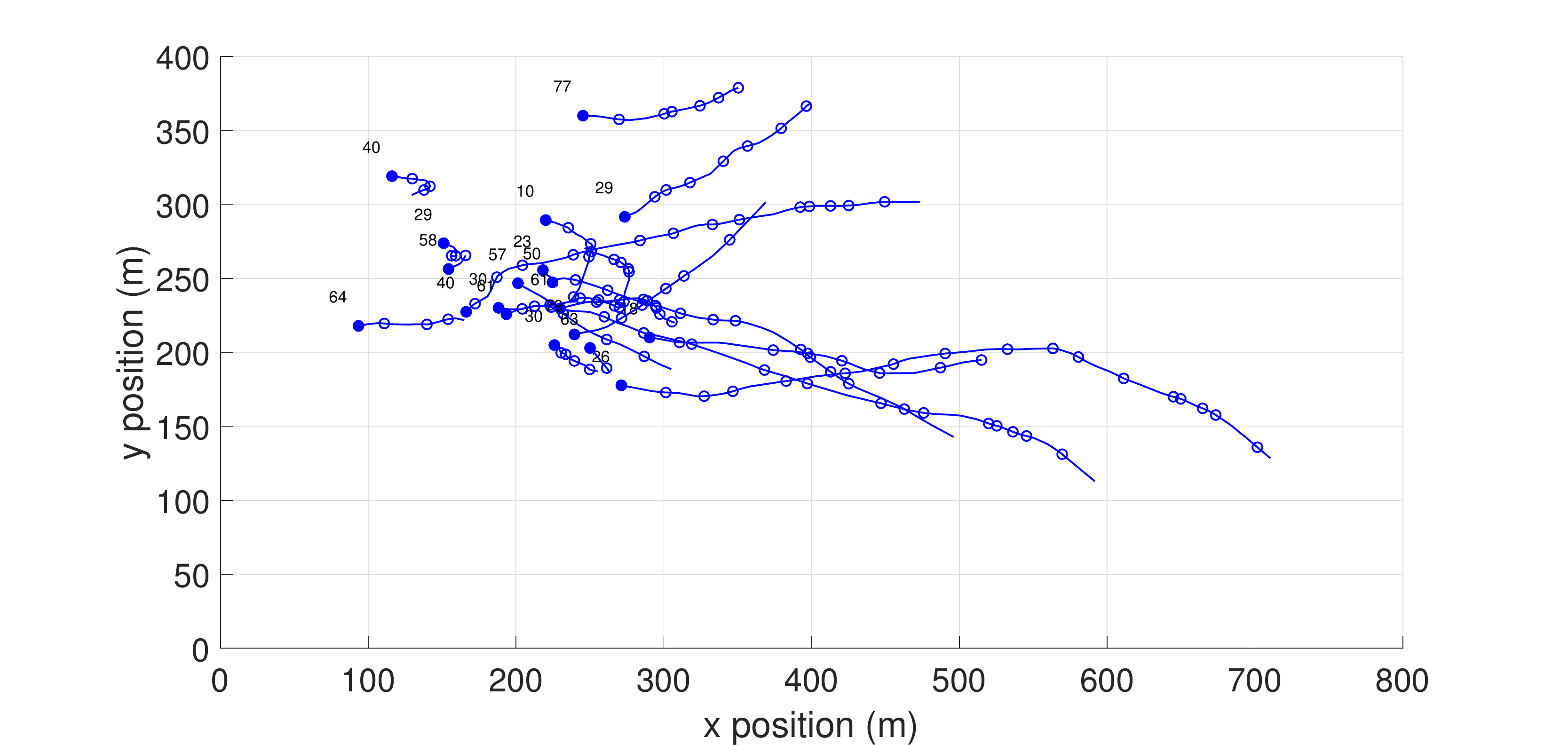}
\par\end{centering}
\caption{\label{fig:Scenario-simulations}Scenario of the simulations with
the set of trajectories sampled at the in-sequence measurement time
steps. The beginning of a trajectory is marked with a filled circle
and its position every 10 time steps is marked with a circle. A number
next to the trajectory start time indicates the (in-sequence) time
step when it was born. }
\end{figure}

The scenario of the simulations is shown in Figure \ref{fig:Scenario-simulations}.
There are 19 targets in total and the maximum number of targets alive
at the same time step is 10. We evaluate the filters via Monte Carlo
simulation with $N_{mc}=100$ runs. For each received measurement
and Monte Carlo run $i$, we calculate the error between the true
set $\mathbf{X}_{k}$ of all trajectories up to the current time and
its estimate $\mathbf{\hat{X}}_{k}^{i}$ (both sampled at in-sequence
sampling times). The error is calculated by the metric $d\left(\cdot,\cdot\right)$
for sets of trajectories in \cite{Angel20_d} with parameters $p=2$,
$c=10$ and $\gamma=1$. We only use the position elements of the
trajectories to compute $d\left(\cdot,\cdot\right)$ and normalise
the squared error by the length of the time window to obtain $d^{2}\left(\mathbf{X}_{k},\mathbf{\hat{X}}_{k}^{i}\right)/k$.
The root mean square (RMS) error at time step $k$ is
\begin{align}
d\left(k\right) & =\sqrt{\frac{1}{N_{mc}k}\sum_{i=1}^{N_{mc}}d^{2}\left(\mathbf{X}_{k},\mathbf{\hat{X}}_{k}^{i}\right)}.\label{eq:error_time_k}
\end{align}

\begin{figure}
\begin{centering}
\includegraphics[scale=0.6]{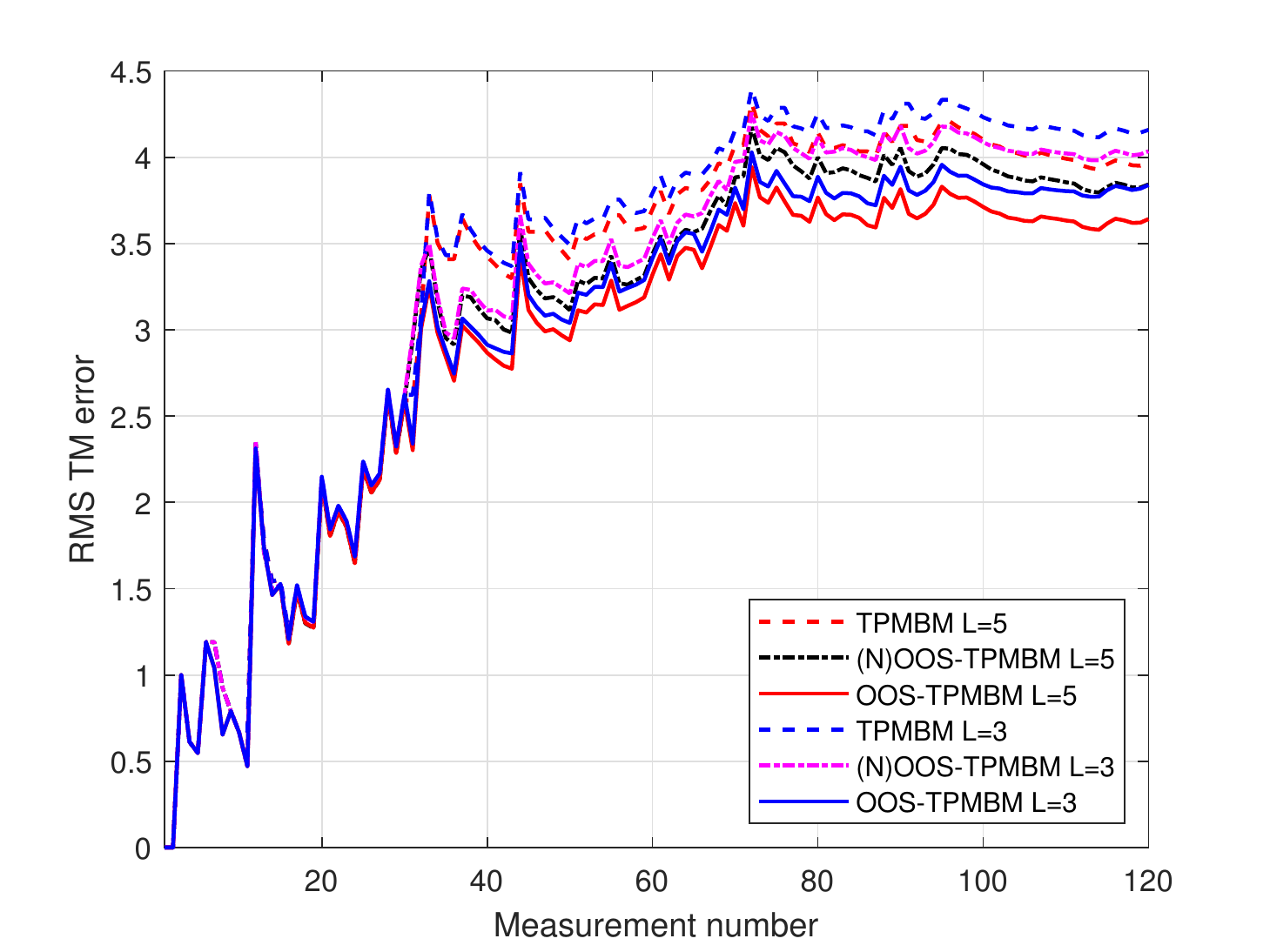}
\par\end{centering}
\caption{\label{fig:RMS-trajectory-metric-error}RMS trajectory metric error
to estimate the set of all trajectories for each received measurement.
The TPMBM filter with optimal OOS processing with $L=5$ has the lowest
error.}
\end{figure}

The RMS trajectory metric (TM) errors (\ref{eq:error_time_k}) of
the TPMBM algorithms against the measurement number are shown in Figure
\ref{fig:RMS-trajectory-metric-error}. As expected, for a given $L$,
the OOS-TPMBM filter is the one with lowest error, followed by the
(N)OOS-TPMBM filter, and the TPMBM filter without OOS processing.
The filters with $L=5$ have lower error than the filters with $L=3$,
as they update a longer time window. We can also see that all filters
have quite similar performance up to around processing 30 measurements,
when differences arise. The reason is that for the first two OOS measurement,
see Figure \ref{fig:Time-difference-measurements}, there are not
any targets present yet, and the processing of the OOS measurements
does not improve performance. It is the processing of the subsequent
OOS measurement that have an impact on performance.

\begin{figure}
\begin{centering}
\includegraphics[scale=0.3]{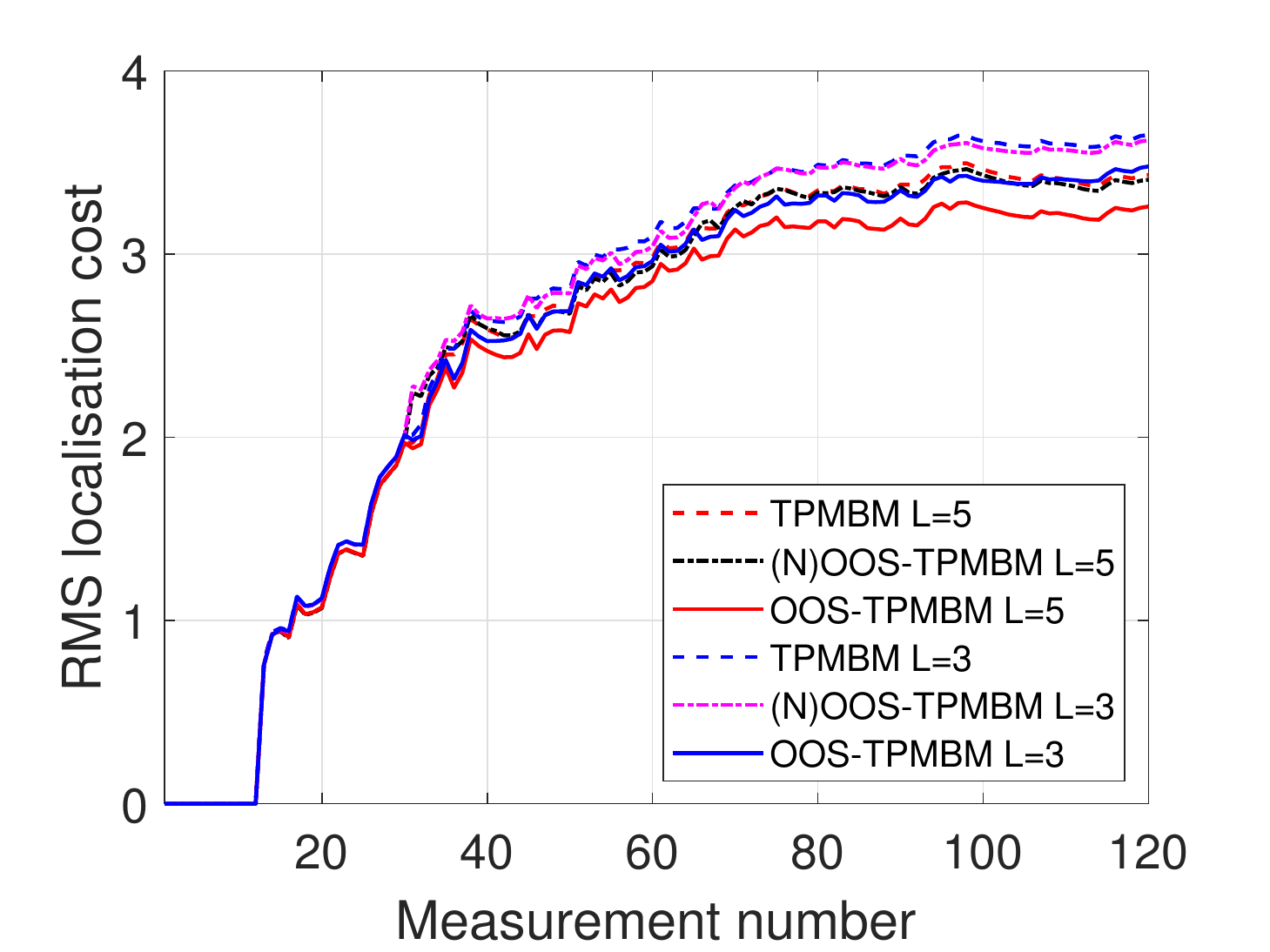}\includegraphics[scale=0.3]{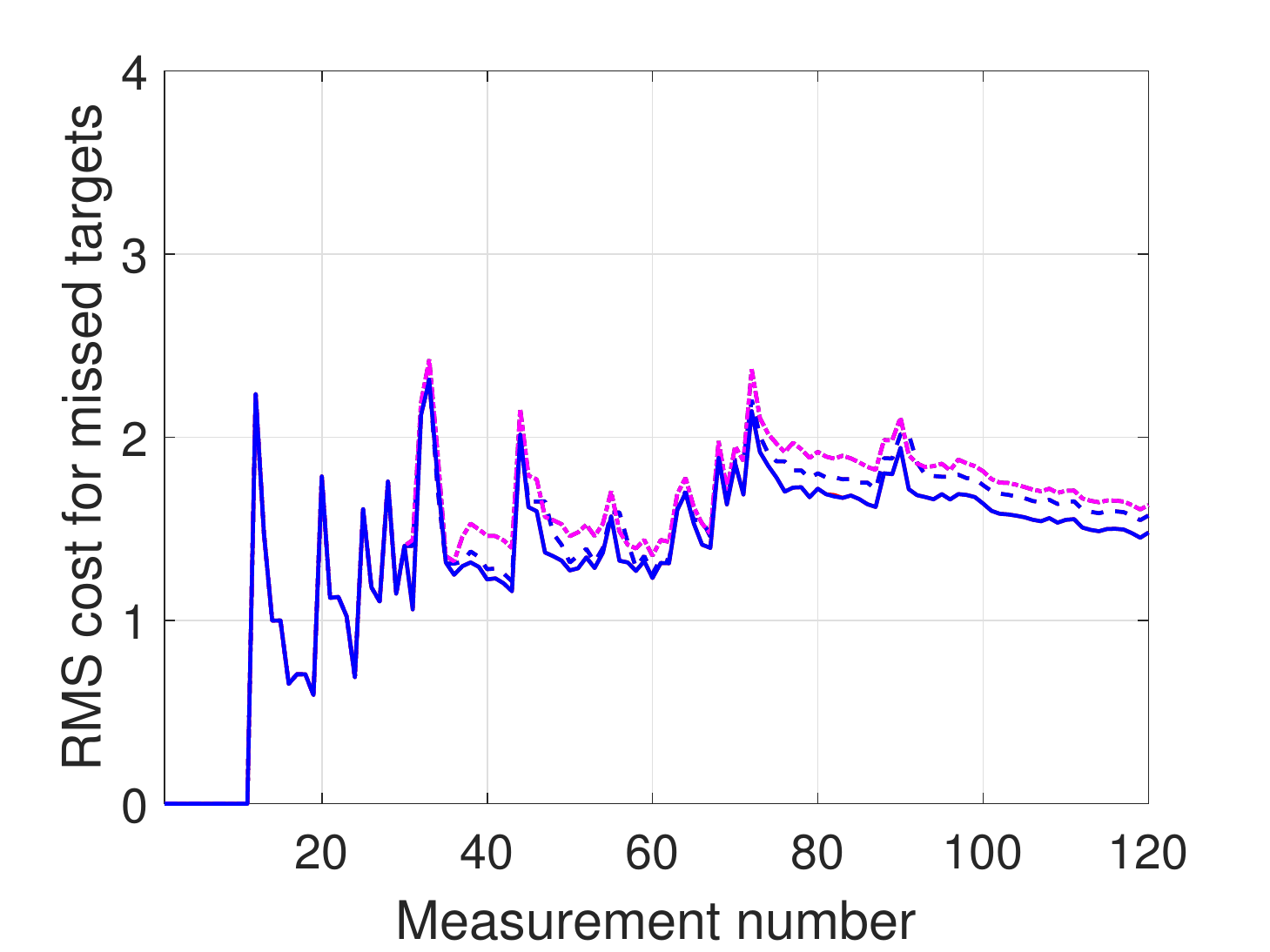}
\par\end{centering}
\begin{centering}
\includegraphics[scale=0.3]{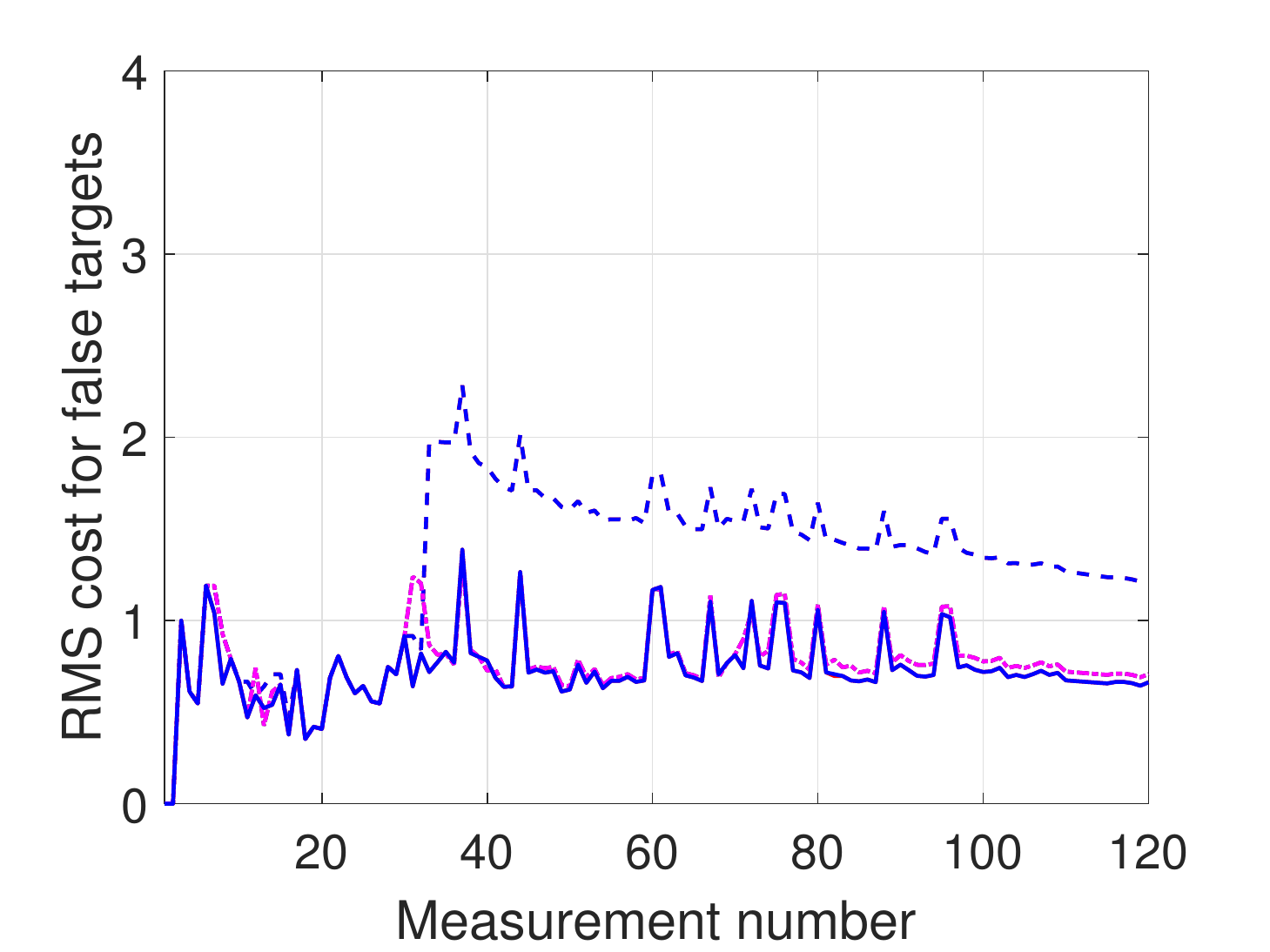}\includegraphics[scale=0.3]{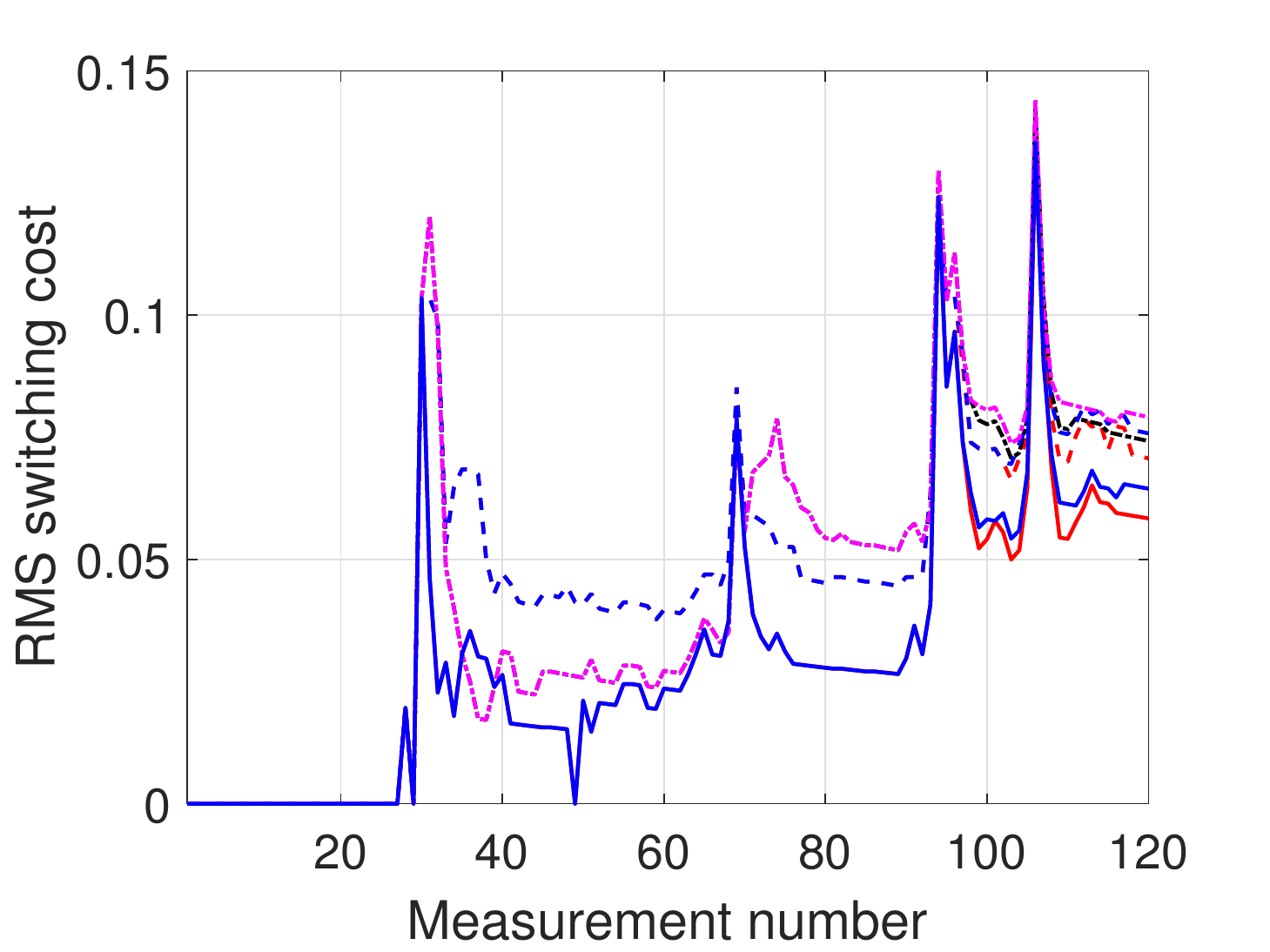}
\par\end{centering}
\caption{\label{fig:Trajectory_metric_decomposition}RMS trajectory metric
decomposition into localisation cost, missed target cost, false target
cost and track switching cost for each received measurement. Filters
with OOS measurement processing mainly lower the cost for false targets
in this scenario. }
\end{figure}

To analyse more thoroughly filter performance, we show the decomposition
of the trajectory metric in Figure \ref{fig:Trajectory_metric_decomposition}.
The filters without OOS processing have a higher false target cost.
The main reason is that the start time of a trajectory (the one born
at at time step 29 with position $\left[151,174\right]^{T}\,\left(\mathrm{m}\right)$)
is estimated more accurately by processing the third OOS measurement.
The filters with optimal OOS processing show better performance than
(N)OOS processing mainly due to improvement in localisation cost.
Increasing $L$ decreases the localisation costs, as the filters
are able to improve estimation of past states. Track switching costs
are zero up to measurement number 26. 

In Table \ref{tab:averaged_RMS}, we show the RMS trajectory metric
error across all time steps \cite{Angel20_e}, also including TPMB
filter performance and the average time to run one Monte Carlo iteration
of our implementations with a 1.6 GHz Intel i5 laptop. As indicated
before, the best performing filter is the OOS-TPMBM with $L=5$. In
this scenario, the TPMB approximation mainly implies an increase in
the number of missed targets, irrespective of the type of OOS processing.
As expected, processing OOS measurements increases running times.
TPMBM filters have higher computational complexity than TPMB filters.
There is little difference in computational times between $L=3$ and
$L=5$.

\begin{table}
\caption{\label{tab:averaged_RMS}RMS trajectory metric and its decomposition
across all time steps, and computational time in seconds}

\centering{}%
\begin{tabular}{cccccccc}
\hline 
$L$ &
Algorithm &
Tot. &
Loc. &
Fal. &
Mis. &
Swi. &
Time\tabularnewline
\hline 
 &
TPMBM &
3.44 &
2.76 &
1.37 &
1.52 &
0.06 &
11.3\tabularnewline
 &
(N)OOS-TPMBM &
3.28 &
2.75 &
0.79 &
1.59 &
0.05 &
11.9\tabularnewline
5 &
OOS-TPMBM &
\uline{3.10} &
2.63 &
0.75 &
1.46 &
0.04 &
12.3\tabularnewline
 &
TPMB &
3.84 &
2.77 &
1.39 &
2.28 &
0.07 &
2.5\tabularnewline
 &
(N)OOS-TPMB &
3.51 &
2.71 &
0.86 &
2.05 &
0.07 &
3.0\tabularnewline
 &
OOS-TPMB &
3.42 &
2.63 &
0.78 &
2.04 &
0.05 &
3.0\tabularnewline
\hline 
\multirow{6}{*}{3} &
TPMBM &
3.54 &
2.88 &
1.37 &
1.52 &
0.06 &
11.0\tabularnewline
 & (N)OOS-TPMBM &
3.37 &
2.87 &
0.79 &
1.59 &
0.06 &
11.8\tabularnewline
 & OOS-TPMBM &
3.20 &
2.74 &
0.75 &
1.46 &
0.04 &
12.3\tabularnewline
 & TPMB &
3.93 &
2.88 &
1.40 &
2.28 &
0.07 &
2.5\tabularnewline
 & (N)OOS-TPMB &
3.61 &
2.83 &
0.87 &
2.05 &
0.07 &
3.0\tabularnewline
 & OOS-TPMB &
3.52 &
2.75 &
0.79 &
2.05 &
0.06 &
3.0\tabularnewline
\hline 
\end{tabular}
\end{table}

\section{Conclusions\label{sec:Conclusions}}

This paper has explained how to perform the Bayesian update with out-of-sequence
measurements for multiple target tracking when the multi-target dynamics
are given in continuous time and we compute the posterior of the set
of all sampled trajectories. When processing in-sequence measurements,
the posterior density of the set of all sampled trajectories is a
Poisson multi-Bernoulli mixture. This paper shows that the processing
of out-of-sequence measurements consists of two steps: retrodiction
and update. After performing these two steps, the posterior is also
a Poisson multi-Bernoulli mixture.

The paper also explains the out-of-sequence measurement processing
when we consider a Gaussian implementation of the trajectory Poisson
multi-Bernoulli mixture filter. Simulation results show that lower
error is achieved by optimally processing out-of-sequence measurements.

\bibliographystyle{IEEEtran}
\bibliography{10C__Trabajo_laptop_Mis_articulos_Out-of-sequence_PMBM_Accepted_Referencias}

\begin{thebibliography}{10}
\providecommand{\url}[1]{#1}
\csname url@samestyle\endcsname
\providecommand{\newblock}{\relax}
\providecommand{\bibinfo}[2]{#2}
\providecommand{\BIBentrySTDinterwordspacing}{\spaceskip=0pt\relax}
\providecommand{\BIBentryALTinterwordstretchfactor}{4}
\providecommand{\BIBentryALTinterwordspacing}{\spaceskip=\fontdimen2\font plus
\BIBentryALTinterwordstretchfactor\fontdimen3\font minus
  \fontdimen4\font\relax}
\providecommand{\BIBforeignlanguage}[2]{{%
\expandafter\ifx\csname l@#1\endcsname\relax
\typeout{** WARNING: IEEEtran.bst: No hyphenation pattern has been}%
\typeout{** loaded for the language `#1'. Using the pattern for}%
\typeout{** the default language instead.}%
\else
\language=\csname l@#1\endcsname
\fi
#2}}
\providecommand{\BIBdecl}{\relax}
\BIBdecl

\bibitem{Challa_book11}
S.~Challa, M.~R. Morelande, D.~Musicki, and R.~J. Evans, \emph{Fundamentals of
  Object Tracking}.\hskip 1em plus 0.5em minus 0.4em\relax Cambridge University
  Press, 2011.

\bibitem{Li93}
X.~R. {Li} and Y.~{Bar-Shalom}, ``Design of an interacting multiple model
  algorithm for air traffic control tracking,'' \emph{IEEE Transactions on
  Control Systems Technology}, vol.~1, no.~3, pp. 186--194, Sep. 1993.

\bibitem{Fortin15}
B.~{Fortin}, R.~{Lherbier}, and J.~{Noyer}, ``A model-based joint detection and
  tracking approach for multi-vehicle tracking with lidar sensor,'' \emph{IEEE
  Transactions on Intelligent Transportation Systems}, vol.~16, no.~4, pp.
  1883--1895, 2015.

\bibitem{Koch11}
W.~Koch and F.~Govaers, ``On accumulated state densities with applications to
  out-of-sequence measurement processing,'' \emph{IEEE Transactions on
  Aerospace and Electronic Systems}, vol.~47, no.~4, pp. 2766--2778, 2011.

\bibitem{Bar-Shalom02}
Y.~Bar-Shalom, ``Update with out-of-sequence measurements in tracking: exact
  solution,'' \emph{IEEE Transactions on Aerospace and Electronic Systems},
  vol.~38, no.~3, pp. 769--777, Jul. 2002.

\bibitem{Bar-Shalom04}
Y.~{Bar-Shalom}, {H. Chen}, and M.~{Mallick}, ``One-step solution for the
  multistep out-of-sequence-measurement problem in tracking,'' \emph{IEEE
  Transactions on Aerospace and Electronic Systems}, vol.~40, no.~1, pp.
  27--37, 2004.

\bibitem{Zhang05}
K.~Zhang, X.~R. {Li}, and Y.~Zhu, ``Optimal update with out-of-sequence
  measurements,'' \emph{IEEE Transactions on Signal Processing}, vol.~53,
  no.~6, pp. 1992--2004, June 2005.

\bibitem{Zhang12}
S.~{Zhang} and Y.~{Bar-Shalom}, ``Out-of-sequence measurement processing for
  particle filter: Exact {B}ayesian solution,'' \emph{IEEE Transactions on
  Aerospace and Electronic Systems}, vol.~48, no.~4, pp. 2818--2831, Oct. 2012.

\bibitem{Orton05}
M.~Orton and A.~Marrs, ``Particle filters for tracking with out-of-sequence
  measurements,'' \emph{IEEE Transactions on Aerospace and Electronic Systems},
  vol.~41, no.~2, pp. 693--702, April 2005.

\bibitem{Govaers14}
F.~{Govaers} and W.~{Koch}, ``Generalized solution to smoothing and
  out-of-sequence processing,'' \emph{IEEE Transactions on Aerospace and
  Electronic Systems}, vol.~50, no.~3, pp. 1739--1748, July 2014.

\bibitem{Shen09}
X.~{Shen}, Y.~{Zhu}, E.~{Song}, and Y.~{Luo}, ``Optimal centralized update with
  multiple local out-of-sequence measurements,'' \emph{IEEE Transactions on
  Signal Processing}, vol.~57, no.~4, pp. 1551--1562, 2009.

\bibitem{Orton02b}
M.~Orton and A.~Marrs, ``A {B}ayesian approach to multi-target tracking and
  data fusion with out-of-sequence measurements,'' in \emph{IEE Target
  Tracking: Algorithms and Applications}, 2001, pp. 15/1--15/5 vol.1.

\bibitem{Zhang03}
K.~Zhang, X.~Li, H.~Chen, and M.~Mallick, ``Multi-sensor multi-target tracking
  with out-of-sequence measurements,'' in \emph{Proceedings of the Sixth
  International Conference of Information Fusion}, vol.~1, 2003, pp. 672--679.

\bibitem{Maskell06}
S.~Maskell, R.~G. Everitt, R.~Wright, and M.~Briers, ``Multi-target
  out-of-sequence data association: Tracking using graphical models,''
  \emph{Information Fusion}, vol.~7, no.~4, pp. 434 -- 447, 2006.

\bibitem{Mallick02}
M.~Mallick, J.~Krant, and Y.~Bar-Shalom, ``Multi-sensor multi-target tracking
  using out-of-sequence measurements,'' in \emph{Proceedings of the Fifth
  International Conference on Information Fusion}, vol.~1, 2002, pp. 135--142.

\bibitem{Chan08}
S.~Chan and R.~Paffenroth, ``Out-of-sequence measurement updates for
  multi-hypothesis tracking algorithms,'' in \emph{Proceedings SPIE Signal and
  Data Processing of Small Targets}, vol. 6969, pp. 1--12.

\bibitem{Yang20}
Q.~Yang and W.~Yi, ``An efficient {PHD} filter for multi-target tracking with
  out-of-sequence measurement,'' in \emph{IEEE Radar Conference}, 2020, pp.
  1--6.

\bibitem{Angel20}
A.~F. García-Fernández and S.~Maskell, ``Continuous-discrete multiple target
  filtering: {PMBM}, {PHD} and {CPHD} filter implementations,'' \emph{IEEE
  Transactions on Signal Processing}, vol.~68, pp. 1300--1314, 2020.

\bibitem{Coraluppi14}
S.~Coraluppi and C.~A. Carthel, ``If a tree falls in the woods, it does make a
  sound: multiple-hypothesis tracking with undetected target births,''
  \emph{IEEE Transactions on Aerospace and Electronic Systems}, vol.~50, no.~3,
  pp. 2379--2388, July 2014.

\bibitem{Angel20_f}
A.~F. García-Fernández and S.~Maskell, ``Continuous-discrete trajectory {PHD}
  and {CPHD} filters,'' in \emph{23rd International Conference on Information
  Fusion}, 2020, pp. 1--8.

\bibitem{Kleinrock_book76}
L.~Kleinrock, \emph{Queueing Systems}.\hskip 1em plus 0.5em minus 0.4em\relax
  John Wiley \& Sons, 1976.

\bibitem{Sarkka_book19}
S.~S{ä}rkk{ä} and A.~Solin, \emph{Applied Stochastic Differential
  Equations}.\hskip 1em plus 0.5em minus 0.4em\relax Cambridge University
  Press, 2019.

\bibitem{Mahler_book14}
R.~P.~S. Mahler, \emph{Advances in Statistical Multisource-Multitarget
  Information Fusion}.\hskip 1em plus 0.5em minus 0.4em\relax Artech House,
  2014.

\bibitem{Angel20_b}
A.~F. García-Fernández, L.~Svensson, and M.~R. Morelande, ``Multiple target
  tracking based on sets of trajectories,'' \emph{IEEE Transactions on
  Aerospace and Electronic Systems}, vol.~56, no.~3, pp. 1685--1707, Jun. 2020.

\bibitem{Granstrom18}
K.~Granstr{ö}m, L.~Svensson, Y.~Xia, J.~L. Williams, and A.~F.
  García-Fernández, ``Poisson multi-{B}ernoulli mixture trackers: continuity
  through random finite sets of trajectories,'' in \emph{21st International
  Conference on Information Fusion}, 2018, pp. 973--981.

\bibitem{Granstrom19_prov2}
\BIBentryALTinterwordspacing
K.~Granstr{ö}m, L.~Svensson, Y.~Xia, J.~Williams, and A.~F.
  Garc{í}a-Fern{á}ndez, ``Poisson multi-{B}ernoulli mixtures for sets of
  trajectories,'' 2019. [Online]. Available:
  \url{https://arxiv.org/abs/1912.08718}
\BIBentrySTDinterwordspacing

\bibitem{Angel20_e}
A.~F. García-Fernández, L.~Svensson, J.~L. Williams, Y.~Xia, and
  K.~Granstr{ö}m, ``Trajectory {P}oisson multi-{B}ernoulli filters,''
  \emph{IEEE Transactions on Signal Processing}, vol.~68, pp. 4933--4945, 2020.

\bibitem{Williams15b}
J.~L. Williams, ``Marginal multi-{B}ernoulli filters: {RFS} derivation of
  {MHT}, {JIPDA} and association-based {MeMBer},'' \emph{IEEE Transactions on
  Aerospace and Electronic Systems}, vol.~51, no.~3, pp. 1664--1687, July 2015.

\bibitem{Angel18_b}
A.~F. García-Fernández, J.~L. Williams, K.~Granström, and L.~Svensson,
  ``Poisson multi-{B}ernoulli mixture filter: direct derivation and
  implementation,'' \emph{IEEE Transactions on Aerospace and Electronic
  Systems}, vol.~54, no.~4, pp. 1883--1901, Aug. 2018.

\bibitem{Reid79}
D.~Reid, ``An algorithm for tracking multiple targets,'' \emph{IEEE
  Transactions on Automatic Control}, vol.~24, no.~6, pp. 843--854, Dec. 1979.

\bibitem{Brekke17}
E.~F. Brekke and M.~Chitre, ``The multiple hypothesis tracker derived from
  finite set statistics,'' in \emph{20th International Conference on
  Information Fusion}, 2017, pp. 1--8.

\bibitem{Brekke18}
E.~Brekke and M.~Chitre, ``Relationship between finite set statistics and the
  multiple hypothesis tracker,'' \emph{IEEE Transactions on Aerospace and
  Electronic Systems}, vol.~54, no.~4, pp. 1902--1917, Aug. 2018.

\bibitem{Bostrom-Rost21_early}
P.~{Bostr{ö}m-Rost}, D.~{Axehill}, and G.~{Hendeby}, ``Sensor management for
  search and track using the {P}oisson multi-{B}ernoulli mixture filter,''
  \emph{IEEE Transactions on Aerospace and Electronic Systems}, 2021.

\bibitem{Granstrom20b}
K.~Granstr{ö}m, L.~Svensson, Y.~Xia, A.~F. García-Fernández, and J.~L.
  Williams, ``Spatiotemporal constraints for sets of trajectories with
  applications to {PMBM} densities,'' in \emph{23rd International Conference on
  Information Fusion}, 2020, pp. 1--8.

\bibitem{Frohle20}
M.~{Fröhle}, K.~{Granström}, and H.~{Wymeersch}, ``Decentralized {P}oisson
  multi-{B}ernoulli filtering for vehicle tracking,'' \emph{IEEE Access},
  vol.~8, pp. 126\,414--126\,427, 2020.

\bibitem{Kulkarni_book16}
V.~G. Kulkarni, \emph{Modeling and analysis of stochastic systems}.\hskip 1em
  plus 0.5em minus 0.4em\relax Chapman \& Hall/CRC, 2016.

\bibitem{Xia19_b}
Y.~Xia, K.~Granstr{ö}m, L.~Svensson, A.~F. García-Fernández, and J.~L. Wlliams,
  ``Multi-scan implementation of the trajectory {P}oisson multi-{B}ernoulli
  mixture filter,'' \emph{Journal of Advances in Information Fusion}, vol.~14,
  no.~2, pp. 213--235, Dec. 2019.

\bibitem{Streit_book10}
R.~Streit, \emph{Poisson point processes: Imaging, tracking, and
  sensing}.\hskip 1em plus 0.5em minus 0.4em\relax Springer, 2010.

\bibitem{Crouse17}
D.~F. {Crouse}, ``The tracker component library: free routines for rapid
  prototyping,'' \emph{IEEE Aerospace and Electronic Systems Magazine},
  vol.~32, no.~5, pp. 18--27, 2017.

\bibitem{Angel20_d}
A.~F. García-Fernández, A.~S. Rahmathullah, and L.~Svensson, ``A metric on the
  space of finite sets of trajectories for evaluation of multi-target tracking
  algorithms,'' \emph{IEEE Transactions on Signal Processing}, vol.~68, pp.
  3917--3928, 2020.

\bibitem{Chiu_book13}
S.~N. Chiu, D.~Stoyan, W.~S. Kendall, and J.~Mecke, \emph{Stochastic Geometry
  and its Applications}.\hskip 1em plus 0.5em minus 0.4em\relax John Wiley \&
  Sons, 2013.

\bibitem{Angel19_f}
A.~F. García-Fernández and L.~Svensson, ``Trajectory {PHD} and {CPHD}
  filters,'' \emph{IEEE Transactions on Signal Processing}, vol.~67, no.~22,
  pp. 5702--5714, Nov 2019.

\bibitem{Sarkka_book13}
S.~S{ä}rkk{ä}, \emph{Bayesian Filtering and Smoothing}.\hskip 1em plus 0.5em
  minus 0.4em\relax Cambridge University Press, 2013.

\end{thebibliography}

\cleardoublepage{}

{\LARGE{}Continuous-discrete multiple target tracking with out-of-sequence
measurements: Supplementary material}{\LARGE\par}

\appendices{}

\section{\label{sec:Appendix_A}}

In this appendix, we explain how to calculate $p_{1}^{S,o}$ and $p_{2}^{S,o}$
in Proposition \ref{prop:Transition_density}, and also provide the
proof of Theorem \ref{thm:Retrodiction-PMBM}.

\subsection{Probability $p_{1}^{S,o}$}

The probability $p_{1}^{S,o}$ corresponds to the probability that
a trajectory that is alive at time step $k^{o}-1$, but not at $k^{o}$,
is alive at time $\tau$. In this subsection, we denote the time lag
of disappearance of this trajectory w.r.t. $t_{k^{o}-1}$ as $t$,
which implies that the trajectory disappears at time $t+t_{k^{o}-1}$.
Given that the trajectory is alive at time step $k^{o}-1$, the distribution
of the time lag of disappearance is an exponential distribution with
parameter $\mu$ \cite{Angel20}. Then, given that the trajectory
is alive at time step $k^{o}-1$ and dead at time step $k^{o}$, the
distribution of $t$ is a truncated exponential distribution
\begin{align}
p_{k}^{o}\left(t\right) & =\frac{\mu}{1-e^{-\mu\Delta t_{k^{o}}}}e^{-\mu t}\chi_{\left[0,\Delta t_{k^{o}}\right)}\left(t\right).\label{eq:density_time_lag_oos}
\end{align}

Therefore, the probability that this trajectory is alive at time $\tau$
is the probability that $t\geq\Delta t_{o,1}$ (i.e. it disappears
after time $\tau$), which is calculated as
\begin{align}
p_{1}^{S,o} & =\int_{\Delta t_{o,1}}^{\Delta t_{k^{o}}}p_{k}^{o}\left(t\right)dt=p_{k}^{S,o}\left(\Delta t_{o,1}\right)
\end{align}
where $p_{k}^{S,o}\left(\cdot\right)$ is given by (\ref{eq:p_s_OOS}). 

\subsection{Probability $p_{2}^{S,o}$}

The probability $p_{2}^{S,o}$ corresponds to the probability that
a trajectory that is not alive at time step $k^{o}-1$, but it is
at $k^{o}$, is alive at time $\tau$. This implies that this trajectory
is born at time step $k^{o}$ and has appeared between times $t_{k^{o}-1}$
and $t_{k^{o}}$. In this section, we denote the time lag of appearance
w.r.t. $t_{k^{o}}$ as $t$, which implies that the trajectory appears
at time $t_{k^{o}}-t$. The distribution of $t$ is the truncated
exponential distribution (\ref{eq:density_time_lag_oos}) \cite{Angel20}.
Then, the probability that the trajectory is alive at time $\tau$
is the probability that $t\geq\Delta t_{o,2}$ (i.e. it appears before
time $\tau$), which can be calculated as
\begin{align}
p_{2}^{S,o} & =\int_{\Delta t_{o,2}}^{\Delta t_{k^{o}}}p_{k}^{o}\left(t\right)dt=p_{k}^{S,o}\left(\Delta t_{o,2}\right).
\end{align}

\subsection{Proof of Theorem \ref{thm:Retrodiction-PMBM}}

We prove Theorem \ref{thm:Retrodiction-PMBM} by noticing that the
retrodiction step corresponds to a PMBM prediction step with a suitable
choice of single-object transition density and PPP intensity for new
born objects \cite{Williams15b,Granstrom18}.

We first obtain the intensity for the set of OOS new trajectories
$\mathbf{N}.$ The trajectories in $\mathbf{N}$ existed at time step
$\tau$, and appeared and disappeared between time steps $k^{o}-1$
and $k^{o}$. That is, these trajectories appeared in an interval
$\Delta t_{o,1}$ and are alive at its end, and disappeared in the
following time interval $\Delta t_{o,2}$. Due to the continuous time
multi-target model and the independent increments property of PPPs
\cite[pp. 99]{Mahler_book14}, $\mathbf{N}$ is independent of the
sampled set of trajectories $\mathbf{X}_{k}$. In addition, $\mathbf{N}$
corresponds to a thinning operation on a birth PPP with intensity
(\ref{eq:intensity_birth}) (on an interval $\Delta t_{o,1}$ instead
of $\Delta t_{k}$). This thinning operation produces a PPP $\lambda_{\tau,k|k}^{B}\left(\cdot\right)$
with the same spatial distribution \cite{Chiu_book13}. Also, the
expected number $w^{B}\left(\Delta t_{o,1},\Delta t_{o,2}\right)$
of OOS new trajectories, which is the integral of $\lambda_{\tau,k|k}^{B}\left(\cdot\right)$,
is the expected number of trajectories that appear in an interval
$\Delta t_{o,1}$ and are alive at its end, which is given by $\frac{\lambda}{\mu}\left(1-e^{-\mu\Delta t_{o,1}}\right)$
\cite{Angel20,Kulkarni_book16}, multiplied by the probability that
a trajectory disappears in an interval $\Delta t_{o,2}$, which is
given by $\left(1-e^{-\mu\Delta t_{o,2}}\right)$, see (\ref{eq:Probability_survival}).
We then obtain the intensity in (\ref{eq:intensity_birth_OOS}) by
adding the information that the trajectories have length one with
a single state at time $\tau$, and are marked with $\beta=-1$. 

Each $X\in\mathbf{X}_{k}$ is transformed with probability one to
$\left(u,Y\right)\in\mathbf{X}_{k}^{\mathrm{a}}$ and transition density
$g_{\tau,k|k}\left(u,Y|X\right)$ in Proposition \ref{prop:Transition_density}.
With these results, we can now apply the PMBM prediction step \cite{Williams15b,Angel18_b,Granstrom18}
to obtain the density of $\mathbf{Y}_{k}=\mathbf{X}_{k}^{\mathrm{a}}\cup\mathbf{N}$,
which yields Theorem \ref{thm:Retrodiction-PMBM}. 

\section{\label{sec:Appendix_B}}

In this appendix, we prove Lemma \ref{lem:Gaussian_OOS_retrodiction}.
In this lemma, we should first note that $p\left(X\right)$ represents
a trajectory with known start time $\overline{\beta}$ and end time
$\omega$, so the trajectory integral (\ref{eq:single_trajectory_integral})
reduces to a standard integral on an Euclidean space. The first entry
in (\ref{eq:Gaussian_OOS_retrodiction}) is the integral w.r.t. a
Dirac delta on the single-trajectory space, which leaves the density
$p\left(\cdot\right)$ unchanged, evaluated at $Y$. The third entry
corresponds to the transition density applied to a density that is
present at time step $k^{o}-1$ but not at $k^{o}$. The output has
two terms. The first one is the integral w.r.t. a Dirac delta that
leaves $p\left(\cdot\right)$ unchanged. The second term is straightforward
as the transition density is a linear/Gaussian dynamic model with
transition matrix $F_{1}$ and covariance matrix $Q_{1}$, which is
extended to include full trajectory information, see also \cite{Angel19_f,Angel20_e}.
The second and fourth entries in (\ref{eq:Gaussian_OOS_retrodiction})
are more complicated, so we analyse them in the next subsections.

\subsection{Trajectory present at $k^{o}-1$ and $k^{o}$\label{subsec:Trajectory-present-present_appendix}}

The second entry corresponds to the transition density applied to
a density of a trajectory that is present at both time steps $k^{o}-1$
and $k^{o}$. We proceed to calculate the corresponding transition
density (\ref{eq:transition_OOS_alive_alive}) for the Wiener velocity
model. For notational simplicity, we denote $x_{1}$ and $x_{2}$
the states of $\left(\overline{\beta},x^{1:i}\right)$ at steps $k^{o}-1$
and $k^{o}$. Then, (\ref{eq:transition_OOS_alive_alive}) is analogous
to the Kalman filter update of a prior
\begin{align*}
 & \mathcal{N}\left(y;F_{1}x_{1},Q_{1}\right)
\end{align*}
with a measurement density (on $x_{2}$)
\begin{align*}
 & \mathcal{N}\left(x_{2};F_{2}y,Q_{2}\right).
\end{align*}

By direct application of the Kalman filter update \cite{Sarkka_book13},
we obtain
\begin{align}
p\left(y|x_{1},x_{2}\right) & =\mathcal{N}\left(y;\left[F_{1}-K_{pp}F_{2}F_{1},\:K_{pp}\right]\left[\begin{array}{c}
x_{1}\\
x_{2}
\end{array}\right],Q_{pp}\right)
\end{align}
where $K_{pp}$ and $Q_{pp}$ are defined in Lemma \ref{lem:Gaussian_OOS_retrodiction}.
Then, the integral of Lemma \ref{lem:Gaussian_OOS_retrodiction} corresponds
to the density of a Gaussian density augmented with another state
$y$ whose conditional density is Gaussian. The result is a Gaussian
with moments in Lemma \ref{lem:Gaussian_OOS_retrodiction}. Note that
in the lemma we write the transition matrix $F_{pp}$ applied to the
whole trajectory, not only to $x_{1}$ and $x_{2}$. 

\subsection{Trajectory present at $k^{o}$ but not at $k^{o}-1$}

The fourth entry corresponds to the transition density applied to
a density of a trajectory that is present at time step $k^{o}$ but
not at $k^{o}-1$. There are two terms in the output. The first one
corresponds to the Dirac delta and leaves the density $p\left(\cdot\right)$
unaltered. It represents that the trajectory appeared at a time between
$\tau$ and $t_{k^{o}}$. The second term considers the hypothesis
that the trajectory appeared at a time between $t_{k^{o}-1}$ and
time $\tau$, and is therefore alive at time $\tau$ and at time step
$t_{k^{o}}$. We proceed to compute this term by first calculating
the transition density (\ref{eq:transition_OOS_dead_alive}) for the
Wiener velocity model. 

The integral w.r.t. $t$ in (\ref{eq:transition_OOS_dead_alive})
is approximated by the Gaussian that minimises the KLD. Its mean and
covariance are denoted by $\overline{x}_{b,1}$ and $P_{b,1}$ and
are given by Prop. 2 in \cite{Angel20} using $\Delta t_{o,1}$ as
the time interval. After this approximation, the transition density
$p\left(y|x^{1}\right)$ can be calculated as the Kalman filter update
of a prior with moments $\overline{x}_{b,1}$ and $P_{b,1}$ and a
measurement density (on $x^{1}$)
\begin{align}
 & \mathcal{N}\left(x^{1};F_{2}y,Q_{2}\right).
\end{align}
This yields
\begin{align}
p\left(y|x^{1}\right) & =\mathcal{N}\left(y;\left(I-K_{np}F_{2}\right)\overline{x}_{b,1}+K_{np}x^{1},Q_{np}\right),
\end{align}
where $K_{np}$ and $Q_{np}$ are given in Lemma \ref{lem:Gaussian_OOS_retrodiction}.
Then, as in Section \ref{subsec:Trajectory-present-present_appendix},
the output of the integral is Gaussian with the moments in Lemma \ref{lem:Gaussian_OOS_retrodiction}.

\section{\label{sec:Appendix_C}}

In this appendix, we provide more details on the TPMBM Gaussian update
\cite{Granstrom18,Angel20_e} with OOS measurements. In particular,
we provide the steps on how to compute the updated local hypotheses
for previous Bernoulli components, which is the main difficulty in
the update.

In the standard TPMBM Gaussian update for in-sequence measurements,
there is only one hypothesis (term in the mixture) of each Bernoulli,
see (\ref{eq:single_trajectory_Gaussian_all}) and \cite[Eq. (64)]{Angel20_e},
that has information on the current state of the trajectory. On the
contrary, for OOS measurement processing, there may be more than one
term that has information on the state at OOS measurement time due
to the application of Lemma \ref{lem:Gaussian_OOS_retrodiction} to
each Gaussian in (\ref{eq:single_trajectory_Gaussian_all}). 

We write the retrodicted single-target density for previous Bernoulli
$i$ with local hypothesis $a^{i}$ as

\begin{align}
 & p_{\tau,k|k}^{i,a^{i}}\left(u,Y\right)\nonumber \\
 & =\delta_{0}\left[u\right]\sum_{\kappa=\beta^{i,a^{i}}}^{k}\alpha_{0,k|k}^{i,a^{i}}\left(\kappa\right)\mathcal{N}\left(Y;\beta^{i,a^{i}},\overline{x}_{k|k}^{i,a^{i}}\left(\kappa\right),P_{k|k}^{i,a^{i}}\left(\kappa\right)\right)\nonumber \\
 & +\delta_{1}\left[u\right]\sum_{\kappa=\beta^{i,a^{i}}}^{k}\alpha_{1,k|k}^{i,a^{i}}\left(\kappa\right)\mathcal{N}\left(Y;\beta^{i,a^{i}},\overline{x}_{\tau,k|k}^{i,a^{i}}\left(\kappa\right),P_{\tau,k|k}^{i,a^{i}}\left(\kappa\right)\right)\label{eq:retrodicted_single_trajectory_Gaussian}
\end{align}
where
\begin{align}
\alpha_{0,k|k}^{i,a^{i}}\left(\kappa\right) & =\alpha_{k|k}^{i,a^{i}}\left(\kappa\right)\left(1-p\left(\kappa\right)\right)\\
\alpha_{1,k|k}^{i,a^{i}}\left(\kappa\right) & =\alpha_{k|k}^{i,a^{i}}\left(\kappa\right)p\left(\kappa\right)
\end{align}
where $p\left(\kappa\right)\in\left\{ 0,1,p_{1}^{S,o},p_{2}^{S,o}\right\} $
depending on the corresponding entry of Lemma \ref{lem:Gaussian_OOS_retrodiction}
for each Gaussian in (\ref{eq:single_trajectory_Gaussian_all}). The
Gaussian components that are augmented with an OOS state have $u=1$,
mean and covariances $\overline{x}_{\tau,k|k}^{i,a^{i}}\left(\kappa\right)$
and $P_{\tau,k|k}^{i,a^{i}}\left(\kappa\right)$, and are included
in the third line in (\ref{eq:retrodicted_single_trajectory_Gaussian}).
The Gaussian components without state augmentation have $u=0$, remain
unchanged w.r.t. the prior, and are included in the second line in
(\ref{eq:retrodicted_single_trajectory_Gaussian}).

In the rest of the appendix, Sections \ref{subsec:Misdetection-hypothesis}
and \ref{subsec:Detection-hypothesis} explain the update with a misdetection
and a detection, respectively. 

\subsection{Misdetection hypothesis\label{subsec:Misdetection-hypothesis}}

The update with misdetection hypothesis of a Bernoulli with single-trajectory
density (\ref{eq:retrodicted_single_trajectory_Gaussian}) is
\begin{align}
 & p_{\tau,k|\tau,k}^{i,a^{i}}\left(u,Y\right)\nonumber \\
 & \propto\delta_{0}\left[u\right]\sum_{\kappa=\beta^{i,a^{i}}}^{k}\alpha_{0,k|k}^{i,a^{i}}\left(\kappa\right)\mathcal{N}\left(Y;\beta^{i,a^{i}},\overline{x}_{k|k}^{i,a^{i}}\left(\kappa\right),P_{k|k}^{i,a^{i}}\left(\kappa\right)\right)\nonumber \\
 & +\delta_{1}\left[u\right]\sum_{\kappa=\beta^{i,a^{i}}}^{k}\alpha_{1,k|\tau,k}^{i,a^{i}}\left(\kappa\right)\mathcal{N}\left(Y;\beta^{i,a^{i}},\overline{x}_{\tau,k|k}^{i,a^{i}}\left(\kappa\right),P_{\tau,k|k}^{i,a^{i}}\left(\kappa\right)\right)\label{eq:misdetection_OOS_update}
\end{align}
where, in this case,
\begin{align}
\alpha_{1,k|\tau,k}^{i,a^{i}}\left(\kappa\right) & =\left(1-p^{D}\right)\alpha_{1,k|k}^{i,a^{i}}\left(\kappa\right).\label{eq:misdetection_OOS_update_weight}
\end{align}

We can see that the Gaussian densities in (\ref{eq:misdetection_OOS_update})
are unchanged after the update. The update only changes the weights
of terms with an OOS state by multiplying them by $\left(1-p^{D}\right)$,
see (\ref{eq:misdetection_OOS_update_weight}). 

The weight and existence probability of the updated Bernoulli with
local hypothesis $a^{i}$ become \cite{Williams15b,Granstrom18}
\begin{align}
w_{\tau,k|\tau,k}^{i,a^{i}} & =w_{k|k}^{i,a^{i}}\left(1-r_{k|k}^{i,a^{i}}p^{D}\sum_{\kappa=\beta^{i,a^{i}}}^{k}\alpha_{1,k|k}^{i,a^{i}}\left(\kappa\right)\right)\\
r_{\tau,k|\tau,k}^{i,a^{i}} & =\frac{r_{k|k}^{i,a^{i}}\left(1-p^{D}\sum_{\kappa=\beta^{i,a^{i}}}^{k}\alpha_{1,k|k}^{i,a^{i}}\left(\kappa\right)\right)}{1-r_{k|k}^{i,a^{i}}p^{D}\sum_{\kappa=\beta^{i,a^{i}}}^{k}\alpha_{1,k|k}^{i,a^{i}}\left(\kappa\right)}.
\end{align}
It should be noted that $p^{D}\sum_{\kappa=\beta^{i,a^{i}}}^{k}\alpha_{1,k|k}^{i,a^{i}}\left(\kappa\right)$
is the average probability of detection at OOS time for a prior (\ref{eq:retrodicted_single_trajectory_Gaussian}). 

\subsection{Detection hypothesis\label{subsec:Detection-hypothesis}}

The update of a Bernoulli with a single-trajectory density (\ref{eq:retrodicted_single_trajectory_Gaussian})
with a measurement $z$ (corresponding to an updated local hypothesis
$\widetilde{a}^{i}$) is \cite{Williams15b}
\begin{align}
p_{\tau,k|\tau,k}^{i,\widetilde{a}^{i}}\left(u,Y\right) & \propto\delta_{1}\left[u\right]\sum_{\kappa=\beta^{i,a^{i}}}^{k}\alpha_{1,k|\tau,k}^{i,\widetilde{a}^{i}}\left(\kappa\right)\nonumber \\
 & \,\times\mathcal{N}\left(Y;\beta^{i,a^{i}},\overline{x}_{\tau,k|\tau,k}^{i,\widetilde{a}^{i}}\left(\kappa\right),P_{\tau,k|\tau,k}^{i,\widetilde{a}^{i}}\left(\kappa\right)\right)\label{eq:update_Gaussian_append}
\end{align}
where $\overline{x}_{\tau,k|\tau,k}^{i,\widetilde{a}^{i}}\left(\kappa\right)$
and $P_{\tau,k|\tau,k}^{i,\widetilde{a}^{i}}\left(\kappa\right)$
are obtained by a Kalman filter update on a Gaussian single-trajectory
prior with mean $\overline{x}_{\tau,k|k}^{i,a^{i}}\left(\kappa\right)$
and covariance $P_{\tau,k|k}^{i,a^{i}}\left(\kappa\right)$, see (53)-(57)
in \cite{Angel20_e}. In addition, the weights of the mixture in (\ref{eq:update_Gaussian_append})
are 
\begin{align}
\alpha_{1,k|\tau,k}^{i,\widetilde{a}^{i}}\left(\kappa\right) & =p^{D}\mathcal{N}\left(z;\overline{z}^{i,a^{i}}\left(\kappa\right),S^{i,a^{i}}\left(\kappa\right)\right)\alpha_{1,k|k}^{i,a^{i}}\left(\kappa\right)\label{eq:posterior_weigth_append}
\end{align}
where $\overline{z}^{i,a^{i}}\left(\kappa\right)$ and $S^{i,a^{i}}\left(\kappa\right)$
are the mean and covariance matrix of the predicted measurement for
hypothesis $\kappa$ \cite{Angel20_e}. 

The weight and existence probability of the updated Bernoulli with
local hypothesis $\widetilde{a}^{i}$ are
\begin{align}
w_{\tau,k|\tau,k}^{i,\widetilde{a}^{i}} & =w_{k|k}^{i,a^{i}}r_{k|k}^{i,a^{i}}\sum_{\kappa=\beta^{i,a^{i}}}^{k}\alpha_{1,k|\tau,k}^{i,a^{i}}\left(\kappa\right)\\
r_{\tau,k|\tau,k}^{i,\widetilde{a}^{i}} & =1.
\end{align}
As this is a detection hypothesis, we have that the updated probability
$r_{\tau,k|\tau,k}^{i,\widetilde{a}^{i}}$ of existence is 1. Moreover,
the Gaussian components with factor $\delta_{0}\left[u\right]$ in
the prior (\ref{eq:retrodicted_single_trajectory_Gaussian}) cannot
be detected, as they do not exist at OOS time, so they do not appear
in the posterior (\ref{eq:update_Gaussian_append}). Then, the posterior
weight $\alpha_{1,k|\tau,k}^{i,\widetilde{a}^{i}}\left(\kappa\right)$,
see (\ref{eq:posterior_weigth_append}), depends on its previous weight
$\alpha_{1,k|k}^{i,a^{i}}\left(\kappa\right)$ and how well this component
explains the received measurement.
\end{document}